\documentclass[preprint,12pt]{elsarticle}

\usepackage{amssymb}
\usepackage{amsmath}
\usepackage{graphicx}

\usepackage{natbib}

\usepackage{booktabs}

\usepackage{array}

\usepackage{makecell}

\usepackage{tabularx}

\usepackage{geometry}  
\usepackage{multirow} 

\usepackage{ragged2e}
\usepackage{subcaption}
\usepackage{multirow}
\usepackage{float}
\usepackage{siunitx}
\usepackage{longtable} 
\usepackage{enumitem}

\usepackage{adjustbox}
\usepackage{url}
\usepackage{tabularray}
\UseTblrLibrary{booktabs}
\usepackage{xurl}
\usepackage[hidelinks]{hyperref}
\usepackage{enumitem} 

\journal{Elsevier}

\begin{document}

\begin{frontmatter}



\title{Fast Frequency Services from HVDC-Connected Offshore Wind Power Plants: A Review in the European Context}


\author{Zhenghua Xu\textsuperscript{a,*}, George Alin Raducu\textsuperscript{a}, Behnam Nouri\textsuperscript{b}, Oscar Saborío-Romano\textsuperscript{c}, Nicolaos A. Cutululis\textsuperscript{c}} 

\affiliation{organization={Vattenfall Vindkraft A/S},
            addressline={Jupitervej 6}, 
            city={Kolding},
            postcode={6000}, 
            country={Denmark}}

\affiliation{organization={Vattenfall Winkraft Europe GmbH},
            addressline={Amerigo-Vespucci-Platz 2}, 
            city={Hamburg},
            postcode={20457}, 
            country={Germany}}
            
\affiliation{organization={Department of Wind and Energy Systems, Technical University of Denmark},
            addressline={Frederiksborgvej 399}, 
            city={Roskilde},
            postcode={4000}, 
            country={Denmark}}

\cortext[1]{Corresponding author: Zhenghau Xu\\
E-mail: zhenghua.xu@vattenfall.com; zhexu@dtu.dk
\\ ORCID: https://orcid.org/0000-0003-3037-4479}

\begin{abstract}
The rapid expansion of offshore wind energy is central to the European Union’s climate-neutrality targets, with High Voltage Direct Current-connected offshore wind power plants (HVDC-OWPPs) becoming increasingly important for integrating gigawatt-scale renewable generation over long distances. Simultaneously, the phase-out of conventional synchronous generators, together with the gradual withdrawal of renewable-energy subsidies, are creating growing demand for ancillary service provision from HVDC-OWPPs as alternatives. This paper presents a comprehensive review of fast frequency services from HVDC-OWPPs in the European context, focusing on inertia support, fast frequency reserve (FFR), and frequency containment reserve (FCR). The review covers both mandatory grid-code requirements and commercial service products, including the latest developments from ENTSO-E, ACER, NESO, TenneT, and Energinet. Furthermore, the paper discusses the key technical gaps and challenges in the practical implementation of fast frequency services from HVDC-OWPPs, highlighting the limited fast response capacity, complexity of stacking optimization, asymmetric provision of inertia, and overall coordination of an HVDC-OWPP system.

\end{abstract}

\begin{graphicalabstract}
\end{graphicalabstract}

\begin{highlights}
\item Focus on the emerging services of inertia, fast frequency reserve, and frequency containment reserve
\item Cover both grid-code obligations and commercial service products
\item Consider not only adopted regulations, but also ongoing development documents and consultation materials
\item Identify the technical gaps and challenges in practical implementation details
\end{highlights}


\end{frontmatter}


\section{Introduction} \label{sec:intro}

\subsection{Context}
In an effort to be climate neutral by 2050, the European Union (EU) has issued a new set of directives aimed at escalating renewable energy generation and increasing its share in total energy consumption to 42.5\% by 2030 (up from 22\% in 2020) \cite{noauthor_renewable_nodate,noauthor_directive_2023}, where the deployment of offshore wind energy generation is the core \cite{noauthor_offshore_nodate,noauthor_communication_2023}, with large-scale projects being underway, such as the energy islands in the North Sea (10 gigawatt (GW)) \cite{noauthor_north_nodate} and at Bornholm (3 GW) \cite{noauthor_energy_nodate}, etc. 

As offshore wind power development is moving towards large capacity and long-distance transmission, High Voltage Direct Current (HVDC) transmission is standing out to integrate the GW-scale offshore wind power into onshore power systems over more than one hundred kilometers, presenting major advantages in both transmission capacity and overall cost-effectiveness \cite{yang_critical_2022,wu2024grid}. 


Before the ongoing energy transition, at the transmission system level, ancillary services were largely provided by synchronous generators powered by fossil-fuels, playing a pivotal role in maintaining power system stability. But as the energy mix for electricity generation is shifting and fossil-fuels are being phased out, there would be less synchronous generators available for providing the services, necessitating alternative sources to fulfill the provision \cite{rancilio_ancillary_2022}. Given the role as a large-scale clean power supply, HVDC-connected offshore wind power plants (HVDC-OWPPs) are thus expected to participate in ancillary service delivery and supporting power system stability. At the same time, financial and policy incentives for OWPPs, such as subsidies \cite{erraia_offshore_2023} and priority in dispatch \cite{noauthor_commission_2020}, etc., are fading away. Instead, revenue streams from ancillary service markets may become a crucial driver for the further expansion of offshore wind power generation \cite{noauthor_new_2018,jansen_offshore_2020}.

\begin{figure}[htbp]
    \centering
    \includegraphics[width=0.8\textwidth]{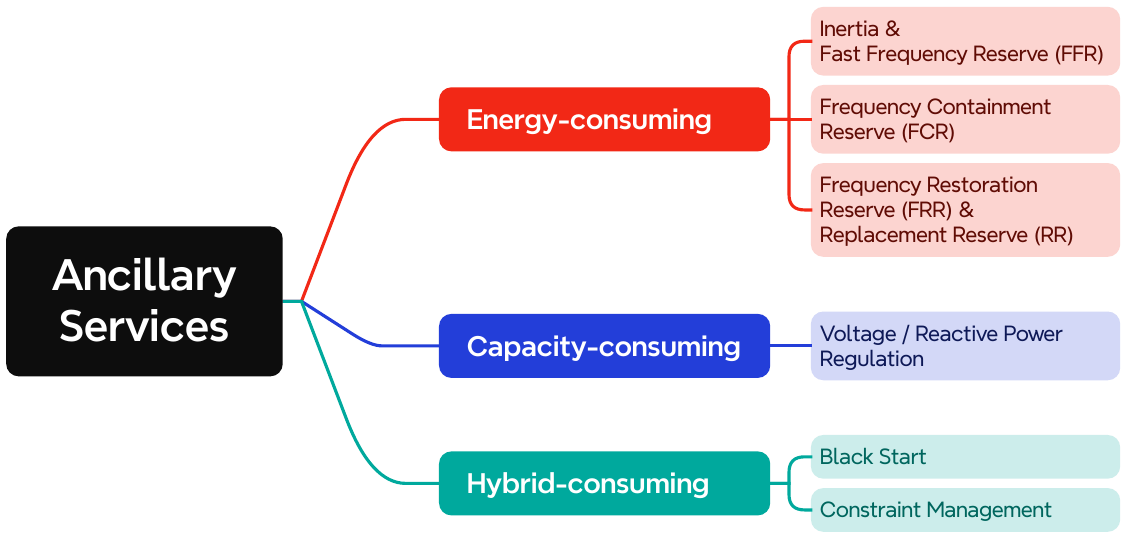} 
    \caption{Existing ancillary services at the transmission system level.}
    \label{fig:categorization}
\end{figure}

The existing ancillary services at the transmission system level are summarized in Figure~\ref{fig:categorization}, where the services are classified according to the resources consumed: energy and converter capacity. Energy-consuming services mainly require active power provision, coming from primary energy and/or stored energy (kinetic, electrochemical, etc.), while capacity-consuming services rely primarily on the capability of the power converters. Hybrid-consuming services require both resources.

At present, OWPPs primarily participate in the energy markets of manual and automatic frequency restoration reserve (mFRR \& aFRR), as these markets operate on intraday timescales (Gate closure time: 25 minutes before delivery) with separate upward and downward products and relatively short delivery periods (15 min) \cite{NextKraftwerke_mFRR,NextKraftwerke_aFRR}. This structure enables OWPPs to participate flexibly in downward regulation only, benefiting from no curtailment, better forecast accuracy, and consequently lower opportunity cost and operational risk.

Participation in other ancillary service markets remains limited. First, existing reactive power markets are currently focused on onshore grid support \cite{netztransparenz_reactive_power_12h_enwg} and OWPPs' reactive power cannot be transported directly to onshore grids through the HVDC links. Consequently, only the onshore HVDC station may participate in present reactive power markets (e.g., dogger bank C in the UK \cite{sse_dogger_bank_reactive_power_2022}). Second, although market-based procurement mechanisms for black-start capability have emerged \cite{netztransparenz_blackstart_market_procurement}, field demonstrations using HVDC-OWPPs remain unavailable. The latest relevant progress is using a 69MW onshore wind farm to re-energise part of the power grid \cite{ScottishPower2020GlobalFirst}. Third, congestion management services are typically location-specific and requested by the transmission system operator (TSO) \cite{neso_transmission_constraint_management,netztransparenz_redispatch}, and fully open market mechanisms for such services are not yet widely established. 

Given their fast response capability enabled by power electronic converters \cite{10158921,10130020}, HVDC-OWPPs are expected to have promising future market opportunities in fast frequency support services, including inertia, fast frequency reserve (FFR) and frequency containment reserve (FCR).

\subsection{Latest Development}
As the energy transition progresses, the grid-code obligations and commercial service products associated with the fast frequency services continue to evolve, while new dedicated services are also emerging. 

Regarding the grid-code obligations, for example, the European Union Agency for the Cooperation of Energy Regulators (ACER) submitted to the European Commission its Recommendation on the amendments to the network codes on requirements for grid connection of generators (RfG Regulation) in December 2023 \cite{ACER2023_GCNC_Amendments} and the HVDC Network Code in December 2024 \cite{ACER2024_HVDC_Amendments}. The mandatory requirements related to synthetic inertia, limited frequency sensitive mode (LFSM), and frequency sensitive mode (FSM) are updated in these amendments. European Network of Transmission System Operators for Electricity (ENTSO-E) published Phase II technical report on Grid Forming (GFM) requirements in November 2025 \cite{ENTSOE_GFM_PhaseII_2025}, including synthetic inertia specifications. In Germany, the latest technical requirements for GFM capabilities including synthetic inertia were published in October 2025 \cite{VDEFNN2025GFM}. In Denmark, the last update on mandatory requirements related to LFSM was issued in October 2025 \cite{Energinet2025_NC_RfG_V5}. 

Regarding the commercial service products, for examples, in Great Britain, stability market, a new dedicated market for GFM functions including inertia provision, was introduced in October 2023 \cite{NGESO_2023_stability_midterm_EOI}, while the latest revised GFM guidance note was released in may 2026 \cite{neso_grid_forming_guidance_2026}. In addition, dynamic frequency response services, representing a new suite of fast frequency response products, have been fully implemented since April 2022 \cite{NESO_DynamicServices_DC_DM_DR}, and the broader frequency response framework is currently undergoing further reform \cite{NESO_future_frequency_response}. In Germany, market-based procurement of instantaneous reserve (i.e., inertia) commenced in January 2026 \cite{netztransparenz_inertia_local_grid_stability}. In Denmark, the version 2 report of GFM deployment and the report on pathway to stability market were released in May 2026 \cite{Kwon2026DeploymentGFM,Energinet2026EvidenceStabilityMarket}.

These recent and ongoing updates in the last 36 months indicate that technical requirements and market mechanisms are still evolving, highlighting the need for a timely developer-oriented comparative review of fast frequency services for HVDC-OWPPs in Europe, integrating regulatory requirements, commercial product design, and implementation barriers.

\subsection{Previous Review Works}
Latest reviews on fast frequency service provision from HVDC-OWPPs are summarized in Table.~\ref{tab:sota_summary} and elaborated as follows.
\newcommand{\tabitem}[1]{%
  \par\noindent
  \hangindent=1.2em
  \hangafter=1
  \makebox[1.2em][l]{\textbullet}#1%
}

\begin{table*}[h!]
\centering
\caption{Latest reviews on fast frequency service provision from HVDC-OWPPs}
\label{tab:sota_summary}
\footnotesize
\setlength{\tabcolsep}{3pt}
\renewcommand{\arraystretch}{1.2}
\renewcommand{\cellalign}{tl} 
\begin{tabular}{@{}p{0.11\textwidth}p{0.04\textwidth}p{0.045\textwidth}p{0.18\textwidth}p{0.28\textwidth}p{0.26\textwidth}@{}}
\toprule
\textbf{Topic} & \textbf{Ref.} & \textbf{Year} & \textbf{Scope} & \textbf{Contributions} & \textbf{Limitations} \\
\midrule
\multirow{3}{0.11\textwidth}{Grid Codes}
& \cite{yu_review_2025} & 2025 & DK, DE, UK, IE, ES, CN, IN, US, IEEE, AU, ZA &\tabitem{Definition and capability specification}
  \tabitem{Experiences from TSO \& projects in GFM development}
 & \tabitem{Pre-2023 codes}
 \tabitem{Missing EU developments}
  \tabitem{No activation requirements}\\
& \cite{wu2024grid} & 2024 & DK, DE, UK, IE, EU, TW, JP, NA 
& 
 \tabitem{Definition and capability specification}
  \tabitem{System configuration}
  \tabitem{Control approaches}
  & \tabitem{Pre-2023 codes}
 \tabitem{Missing EU developments}
  \tabitem{No activation requirements}\\
& \cite{li_review_2023} & 2023 & DK, DE, UK, IE, CN, ES, US, CA, AU, IN, ZA, IEEE & 
\tabitem{Definition and capability specification}
 \tabitem{Future trend analysis}
   & \tabitem{Pre-2023 codes}
 \tabitem{Missing EU developments}
  \tabitem{No activation requirements}\\
\midrule
\multirow{3}{0.11\textwidth}{Commercial Products \& Market}
& \cite{ancillarySurvey_vahid} & 2026 & IE, GB, JP & 
 \tabitem{Definitions and classifications}
  \tabitem{Case study in IE, GB, JP}
  \tabitem{Trends and challenges}
&
  \tabitem{Coverage across Europe}
  \tabitem{Technical and market details} \\
& \cite{viola_ancillary_2024} & 2024 & EU \& US ancillary markets & 
  \tabitem{Assess market developments}
  \tabitem{Identifies technical and market-design challenges}
 & 
  \tabitem{Technical and market details}
     \tabitem{Not up to date}\\
& \cite{rancilio_ancillary_2022} & 2022 & EU ancillary market & 
  \tabitem{European development in ancillary market}
  \tabitem{Analyzes market evolution and regulatory trade-offs}
 &  
\tabitem{Technical and market details}
\tabitem{Not up to date}\\
\midrule
\multirow{4}{0.11\textwidth}{Technical Gaps \& Challenges}
& \cite{FreqCtrlReview_Conf,FreqCtrl_review_IEEEAccess} & 2024 2021 & HVDC-OWPP control strategies & 
\tabitem{Comparison of frequency control approaches} &
\tabitem{Implementation details}\\
& \cite{Ullah2024WindStorageFR} & 2024 & Wind + Energy Storage System & 
\tabitem{Synergy of wind and storage}
 & 
 \tabitem{Implementation details}
\tabitem{No discussion on HVDC} \\
& \cite{cole_critical_2023} & 2023 & Modelling \& Control &
\tabitem{Wake \& fatigue modelling}
\tabitem{Turbine \& plant level control}
 &  \tabitem{Implementation details}
\tabitem{No discussion on HVDC}\\
& \cite{Boyle2024FrequencyControlWindIreland} & 2024 & Turbine level control & 
 \tabitem{Synergy with battery}
\tabitem{Implementation details}
 & 
  \tabitem{No discussion on HVDC}
\tabitem{Coverage across Europe} \\
\bottomrule
\end{tabular}
\end{table*}

Regarding the grid-code obligations of fast frequency services from HVDC-OWPPs, recent reviews have not fully captured the latest regulatory developments and have primarily focused on technical capability requirements while overlooking activation requirements. Published in 2024 to 2025, the latest reviews on grid codes for wind power integration \cite{yu_review_2025, wu2024grid} still concentrate on grid codes formalized prior to 2023 and therefore do not reflect the latest updates mentioned above, missing the development of inertia requirements in Europe. These studies review and compare the grid codes for wind power integration across multiple countries, including Denmark, Germany, Great Britain, Ireland, Spain, etc. Although the technical capability requirements related to frequency control are comprehensively addressed, the corresponding activation requirements are not included. Similarly, the review presented in \cite{li_review_2023} in 2023 provides a dedicated survey of grid-code requirements for frequency regulation from wind power plants, focusing on frequency operating range, primary frequency control, and inertia provision. However, it neither incorporates the most recent regulatory developments nor addresses activation requirements 

Regarding commercial fast frequency service products, recent review papers emphasized the breadth of coverage in order to identify overall development trends, whereas detailed analyses of technical requirements and market mechanisms remain limited. A comprehensive survey is provided in \cite{ancillarySurvey_vahid} on the evolving definitions and classifications of ancillary services in modern power systems. Especially, it highlights the emergence of fast frequency response and synthetic inertia. The international case studies covering Ireland, Great Britain, and Japan are examined. However, the associated technical requirements and market mechanisms are introduced only at a high level rather than discussed in detail. In \cite{viola_ancillary_2024}, a systematic review of ancillary service markets in Europe and the United States is presented, aiming to assess the future evolution of modern ancillary service markets and discuss potential solutions to remaining technical and market-design challenges. Nevertheless, the paper does not investigate the detailed implementation of existing technical requirements and market mechanisms, nor does it incorporate the latest market developments. Similarly, in \cite{rancilio_ancillary_2022}, a meta-analysis is conducted on the evolution of ancillary service markets in Europe and the underlying regulatory trade-offs, focusing on trends in market architecture, service design, and market organization, while detailed discussions of technical requirements and market mechanisms are not covered.

Regarding the technical gaps and challenges in HVDC-OWPPs providing fast frequency services, recent reviews focus on control approaches, especially at the turbine level, and the synergy with energy storage system (ESS), with limited attention to the detailed implementation. Reviews in \cite{FreqCtrl_review_IEEEAccess,FreqCtrlReview_Conf} compare various frequency-control strategies for an HVDC-OWPP and emphasize the challenge posed by communication delays in transmitting onshore frequency measurements offshore. The review in \cite{Ullah2024WindStorageFR} covers frequency regulation using wind power and the support from different ESS, highlighting the challenges of the intermittent and unpredictable nature of wind, limited rotor kinetic energy support, and dilemma in applying de-load control, as well as the challenges related to the high ESS cost, sizing complexity, coordination requirements, and profitability concerns. The review in \cite{cole_critical_2023} further reports that advanced frequency-control strategies may increase turbine structural fatigue and that many studies neglect practical factors such as wake effects and turbulence. It further identifies the lack of a unified framework integrating power-system dynamics, wind-farm control, wake interactions, and structural analysis. Review in \cite{Boyle2024FrequencyControlWindIreland} supplements that Schemes based on RoCoF estimation are highly sensitive to sensor noise and measurement errors, while alternative approaches often require accurate wind-speed estimation, detailed turbine models, and precise real-time control, limiting practical applicability.

Therefore, to the best of the authors’ knowledge, there is no in-depth review of the literature focusing on both mandatory and commercial requirements of fast frequency services, simultaneously covering adopted regulations and ongoing developments, and identifying gaps and challenges in their detailed implementation.

\subsection{Scope and Contributions}
 From the perspective of HVDC-OWPP developers, this paper presents a detailed review of fast frequency services in the European context, with the aim of identifying the barriers to exploiting emerging commercial opportunities beyond mandatory requirements. 
 
Accordingly, the scope includes:
\begin{itemize}
\item{\textbf{Services:} Inertia, FFR, FCR }
\item{\textbf{Requirements:} Both grid codes and commercial products}
\item{\textbf{Regulations:} Up-to-date adopted regulations, ongoing development documents, technical proposals, and consultation materials.}
\item{\textbf{Case studies:} General regulatory frameworks established by ENTSO-E and the implementation practices of TSOs associated with operational or emerging HVDC-OWPP integration, including National Energy System Operator (NESO) in Great Britain, TenneT in Germany, and Energinet in Denmark.}
\end{itemize} 
The contributions of this paper include:
\begin{itemize}
\item{\textbf{Developer-oriented perspective:} The review assesses fast frequency services from the perspective of HVDC-OWPP developers, owners, and operators.}
\item{\textbf{Integration of mandatory and commercial requirements:} The review jointly examines grid-code obligations and commercial service specifications to clarify the transition from mandatory compliance to market participation}
\item{\textbf{Forward-looking assessment:} By including regulatory and technical documents still under development, the review captures emerging trends.}
\item{\textbf{Focus on practical implementation:} The review identifies technical gaps, implementation challenges, and insufficiently specified requirements that may hinder service delivery, qualification, and commercialization.}
\end{itemize} 

\begin{figure}[htbp]
    \centering
    \includegraphics[width=\textwidth]{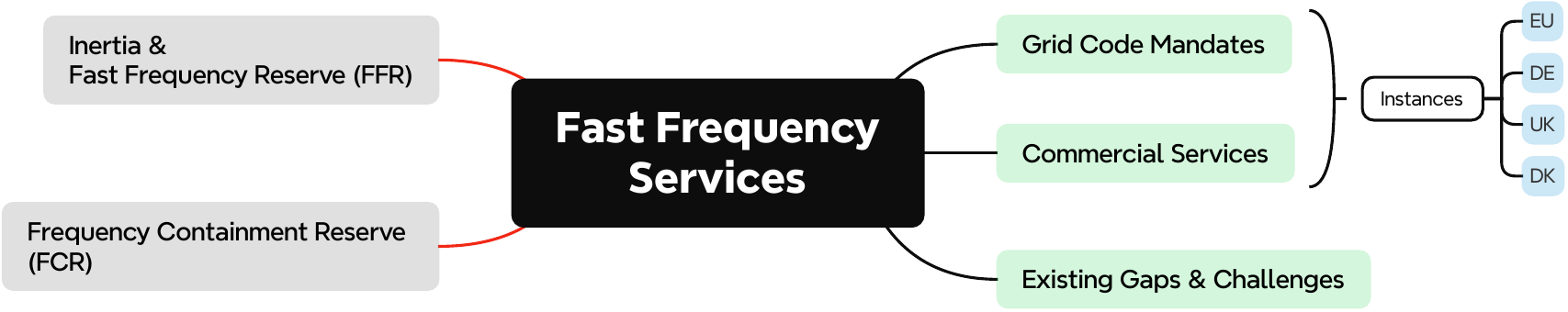} 
    \caption{Review structure.}
    \label{fig:PaperStructure}
\end{figure}

The overall review structure is illustrated in Figure~\ref{fig:PaperStructure}.

The remainder of this paper is structured as follows. Section~\ref{sec:IntaFFR} reviews the grid-code obligations and commercial service products related to inertia and FFR, with the corresponding technical requirements, market mechanisms, and ongoing development summarized and discussed. Section~\ref{sec:FCR} presents a similar review for FCR. Section~\ref{sec:Gaps} identifies the existing gaps and challenges associated with the implementation details of these services by HVDC-OWPPs. Finally, conclusions are drawn in Section~\ref{sec:Conc}.

\section{Inertia and Fast Frequency Reserve} \label{sec:IntaFFR}
Inertia is an active power response proportional to the rate of change of frequency (RoCoF), whereas FFR is a temporary stepwise active power response activated when the frequency reaches a predefined threshold. Both inertia and FFR are ultra-fast, short-duration services designed for low inertia situation to limit transient frequency deviations, although there are nuances in their objectives, as inertia is used to limit the RoCoF after an event until FCR can arrest frequency deviations, and FFR is used to avoid substantial frequency drops resulting from the outage of major generation units or lines. Eventually, inertia and FFR are restored when RoCoF settles to zero or their short delivery period expires. 

In recent development, inertia has been incorporated into GFM capability or service framework, which may also include other functions, such as short-circuit current contribution and island operation, etc. However, since the technical requirements for these functions are generally specified independently, inertia is discussed independently in this section.


\subsection{Grid-code Obligations for Inertia and Fast Frequency Reserve}

\subsubsection{Inertia}

As summarized in Table \ref{tab:IR&FFR_Manda}, generally, the inertia obligations of HVDC-OWPPs are evolving from optional capability toward mandatory capability, accompanied by increasingly comprehensive technical specifications.

\begin{table}[h]
\centering
\resizebox{\columnwidth}{!}{%
\begin{tabular}{|c|c|c|c|c|c|c|}
\hline
\textbf{TSO} & \multicolumn{2}{c|}{\makecell{\textbf{ENTSO-E}\\\textbf{(EU)}}} & \makecell{\textbf{NESO}\\\textbf{(GB)}} & \multicolumn{2}{c|}{\makecell{\textbf{Tennet}\\\textbf{(DE)}}} & \makecell{\textbf{Energinet}\\\textbf{(DK)}} \\
\hline
\textbf{Category} & Inertia & Inertia & \multirow{6}{*}{\makecell{No\\Mandate}} & Inertia & FFR & \multirow{6}{*}{\makecell{No\\Mandate}} \\
\cline{1-3}\cline{5-6}
\textbf{Service} & \makecell{Synthetic Inertia\\(in NC 1.0)}& \makecell{Synthetic Inertia\\(in NC 2.0)} &   & \makecell{Momentanreserve\\(Instantaneous Reserve)} & \makecell{Schnelle\\frequenzgeführte\\ Leistungsregelung\\(Fast Frequency \\Power Regulation)} &   \\
\cline{1-3}\cline{5-6}
\textbf{Target} & Type C \& D & \makecell{Type B: \\Only to high frequency\\ Type C \& D: \\ To high \& low frequency}&   & All & All &   \\
\cline{1-3}\cline{5-6}
\makecell{\textbf{Technical} \\ \textbf{Specifications}} & N/A & \makecell{Mechanical Starting Time\\ defined in \eqref{TmPPM} \\(within inherent capability;\\may require addition\\from Type C \& D) \\ \& Min Damping Ratio \\ during Phase Jump \cite{ENTSOE2025GridFormingPPM}} &   & \makecell{Inertia Time Constant;\\Compliance within \\the Envelope Curves \\based on \\Referenece Behavior} & N/A &   \\
\cline{1-3}\cline{5-6}
\textbf{Requirement} & \makecell{Optional\\Capability} & \makecell{Mandatory\\Capability}&   & \makecell{Mandatory\\Capability} & \makecell{Mandatory\\Capability} &   \\
\cline{1-3}\cline{5-6}
\textbf{Reference} &
\makecell{NC HVDC \cite{noauthor_commission_2016_hvdc}:\\ \text{Art}. 38 \\ NC RfG \cite{noauthor_commission_2016_rfg}: \\\text{Art}. 21(2)} &
\makecell{NC HVDC 2.0 \cite{ACER2024HVDCAmendment}:\\ \text{Art}. 38 \\ NC RfG 2.0 \cite{ACER2023RfGAmendment}: \\\text{Art}. 20(5), 21(5), 22(1)} &  &
\makecell{NAR \cite{NAR}: \\10.2.2, C.2.409\\ VDE 4131 \cite{vde4131}:\\ 10.2.4, 10.1.4.2 }&
\makecell{NAR \cite{NAR}\\10.2.2\\ VDE 4131 \cite{vde4131}:\\ 10.2.4, 10.1.4.3} &
  \\
\hline
\end{tabular}
}
\caption{Grid-code obligations of HVDC-OWPPs for inertia and fast frequency reserve services}
\label{tab:IR&FFR_Manda}
\end{table}

\subparagraph{EU}

At the EU level, the initial inertia obligations in the present Network Codes (NC) require only optional capability from large power plants without any technical specifications, leaving substantial discretion to the relevant TSO. In contrast, the forthcoming Network Codes 2.0 (NC 2.0) will introduce mandatory capability requirements for a broader range of power plants, including both large- and medium-scale ones, together with a methodology (given in \cite{ENTSOE2025GridFormingPPM}) intended to facilitate the development and implementation of detailed technical specifications for the quantification of synthetic inertia provision and its associated dynamic performance. 

The quantification is expressed by mechanical starting time $T_{\text{M,PPM}}$ (s) and inertial response power $\Delta P$ (W) defined as
\begin{subequations}
\label{InertiaDefinition}
\begin{align}
    &T_{\text{M,PPM}} = \frac{\frac{\Delta P}{P_{\text{Rated}}}}{\frac{d(f/f_{\text{Rated}})}{dt}}
    \label{TmPPM}
    \\
    &\Delta P = T_{\text{M,PPM}} \cdot \frac{df/f_{\text{Rated}}}{dt} \cdot P_{\text{Rated}}
    \label{DeltaP}
\end{align}
\end{subequations}
where $P_{\text{Rated}}$ (W) is the rated power; $f$ (Hz) is the power system frequency at the point of connection (POC); $f_{\text{Rated}}$ (Hz) is the rated frequency; a RoCoF ($df/dt$) of 2 Hz/s is typically assumed to evaluate $T_{\text{M,PPM}}$ and $\Delta P$. Commonly, $T_{\text{M,PPM}}$ in \eqref{TmPPM} serves as the main indicator of synthetic inertia provision, following the convention adopted for synchronous machines. Meanwhile, $\Delta P$ in \eqref{DeltaP} provides the direct and practical information on the magnitude of the power response to be delivered. 

While the value of $T_{\text{M,PPM}}$ remains to be specified by the TSO in coordination with the HVDC-OWPP operator, the underlying principle is that the provision should not exceed the inherent capability. Under this principle, no exclusive power headroom is required to be reserved during normal operation. Instead, upward response is required only when headroom exists naturally, e.g. when the HVDC-OWPP is curtailed because the available wind power exceeds the scheduled generation, or when headroom is reserved for the provision of other services such as FCR. Nevertheless, NC 2.0 reserves the right for the TSO to require additional energy beyond the inherent energy storage of large-scale (type C \& D) power plants, subject to coordination with the HVDC-OWPP operator. In such cases, the corresponding power capacity and buffer energy must be ensured at any continuous operating point.

To be technology-agnostic, the dynamic performance evaluation is based on the following metrics:
\begin{itemize}
\item{The expected value of the current or power output}
\item{The response times of the current or power expected value}
\item{The decay rate or overshoot of the current or power excursion (when relevant)}
\item{The damping ratio of the current or power oscillation}
\end{itemize}
To derive the metrics above, the following framework is defined:
\begin{itemize}
\item{Analytical expressions of the expected response}
\item{Compliance test setups based on passive components or grid emulation }
\item{Compliance test cases based on island event and RoCoF event}
\item{Strategy when reaching current capability limit}
\item{Inclusion of additional equipments}
\end{itemize}
However, the inertia provided by PPMs depends on a range of technology-specific factors, including primary energy source characteristics and control algorithms. Besides, the maturity of GFM capabilities and inertia provision in non-storage PPMs is relatively low. Therefore, no specific acceptance criterion is provided, which remains to be provided in the future implementation guidance document (IGD) on GFM capability or the national implementations.

\subparagraph{DE}

In Germany, the regulatory development follows a trajectory similar to that at the EU level. In this regard, Germany provides a representative example of the national-level implementation of the Network Code. Initially, inertia mandates mention only mandatory capability without detailed technical specification, as defined in VDE-AR-N 4131 referred in the main body of Tennet's grid code. Subsequently, the inertia capability was further elaborated in the GFM capability specifications in \cite{VDEFNN2020GFM,VDEFNN2025GFM}. Accordingly, these specifications were adapted and incorporated into the annex of Tennet's grid code, where the quantification and dynamic performance evaluation are detailed.

The German quantification of inertia is still based on $T_{\text{M,PPM}}$ and $\Delta P$ in \eqref{InertiaDefinition}. The value of $T_{\text{M,PPM}}$ is project-specific and defined by the TSO in coordination with the HVDC-OWPP operator. $T_{\text{M,PPM}}$ shall be verified to be maintained at all times. However, temporary deviation is permissible due to converter limits, control priority requirements, or stability constraints. In addition, when a HVDC-OWPP operates at Maximum Power Point Tracking (MPPT), the provision of positive inertia may shift the system toward an operating point that subsequently results in a temporary reduction in feed-in power. In this case, recovery of the internal energy storage after inertia provision by drawing energy from the grid is permitted, provided that the recovery energy does not exceed 1.5 times the energy delivered for inertia provision. 

The German specifications on the dynamic performance of inertia provision are based on the following framework:
\begin{itemize}
\item{Compliance test system based on a Thévenin equivalent for the network}
\item{Compliance test scenarios with disturbances and faults}
\item{Reference behaviour generated by the TSO}
\item{Envelope curves derived from the reference behavior}
\item{Priority of current or power components when reaching the limit}
\item{Inclusion of additional equipment}
\end{itemize}
The test system and scenarios are parameterized in detail. The reference behaviour is generated using a generic device-under-test (DUT) model under an appropriate network equivalent and the defined scenarios. The envelope curves are derived from either the peak and steady-state values or the continuous simulation curve of the reference behaviour, allowing a small margin (0.05 p.u.) for under-provision and a large margin (0.5 p.u.) for over-provision \cite{VDEFNN2020GFM}. Consistent with the EU specifications, no priority shall be attached to any current component, and the inclusion of additional equipments is permitted to support inertia provision, which doesn't increase the capacity base \cite{VDEFNN2025GFM}.

Moreover, during the capability compliance testing, both high and low $T_{\text{M,PPM}}$ (e.g., 20 s and 0.3 s, respectively) are required to be assessed \cite[Annex C.2.409]{NAR}. A low $T_{\text{M,PPM}}$ represents operating conditions where little or no additional energy can be exchanged with the HVDC-OWPP, such as no wind or low rotor speed conditions, whereas a high $T_{\text{M,PPM}}$ corresponds to scenarios where maximum energy can be exchanged.

\subparagraph{GB \& DK}

In Great Britain and Denmark, there is currently no mandate. In particular, the inertia capability is specified as part of the GFM capability in the grid code of Great Britain, but it is not mandated. Instead, the corresponding requirements are applied within the commercial stability market. Therefore, the specifications are discussed in Section~\ref{IR_FFR_commercial} in the context of commercial services.  

\subsubsection{FFR}

At the same time, as shown in Table \ref{tab:IR&FFR_Manda}, FFR is typically not mandated in grid codes but may be required at the discretion of the TSO (e.g., in Germany) based on system-specific studies. Accordingly, the lack of publicly available generic specifications and the project-specific nature of these requirements preclude a general discussion of FFR mandates in this paper.

\subsection{Commercial Products for Inertia and Fast Frequency Reserve}
\label{IR_FFR_commercial}

\subsubsection{Inertia}

Commercial inertia is an emerging ancillary service in Europe. GB pioneered its implementation by establishing a dedicated stability market that includes inertia provision (additional services such as reactive power and restoration are expected to follow, although their procurement will be on a case-by-case basis \cite{NESO_LT2029_ITT_2025}). Subsequently, Germany followed with the introduction of an instantaneous reserve market. Both markets were initially introduced as year-based services, with an overview of the key characteristics summarized in Table~\ref{tab:IR&FFR_Commc}. In addition, there is a stability market under development in Denmark \cite{Kwon2026DeploymentGFM,Energinet2026EvidenceStabilityMarket}.

\begin{table}[h]
    \centering
    \renewcommand{\arraystretch}{1.3}
    \setlength{\tabcolsep}{6pt}

    \resizebox{\textwidth}{!}{
    \begin{tabular}{|c|*{3}{c|}}
        \hline
        \multicolumn{1}{|c|}{} & \textbf{NESO (GB)} 
        & \textbf{Tennet (DE)}
        & \textbf{Energinet (DK)} \\
        \hline  
        \multicolumn{1}{|c|}{\textbf{Category}} & Inertia & Inertia & FFR 
        \\
        \hline
        \textbf{Service} & \makecell{Stability Market} &  Momentanreserve & FFR 
        \\
        \hline
        \textbf{Direction} & \makecell{Symmetric} &  Asymmetric & Upward 
        \\
        \hline
        \textbf{\makecell{Entry\\Conditions}} & \makecell{$\geq$ 132 kV \\ $\geq$ 90\% avail (period unit: 30 min) \\ GFM Capability in Grid Code} & \makecell{GFM Certificate \\for a certain \\Inertia Time Constant}  & 0.3 MW 
        \\
        \hline
        \textbf{Market Size} & \makecell{2025-2026 Mid-term Y-1: 10 GVA$\cdot\text{s}$ \\ 2026-2027 Mid-term Y-1: 15 GVA$\cdot\text{s}$\\ 2027-2028 Mid-term Y-1: 15 GVA$\cdot\text{s}$\\ 2029-2039 Long-term Y-4: 10 GVA$\cdot\text{s}$} &\makecell{Positive: 140 GW$\cdot\text{s}$\\Negative: 259.9 GW$\cdot\text{s}$} & 45 MW 
        \\
        \hline
        \makecell{\textbf{Technical}\\\textbf{Specification}} & \makecell{Inertia Time Constant\\Delay $\leq$ 5 ms \cite[GD.1]{NESO_Grid_Code_2026}\\Symmetrical Inertia Ability \\Due Reactive Power Capability} & \makecell{Inertia Time Constant;\\Compliance within \\the Envelope Curves \\based on \\Referenece Behavior} & \makecell{
Response pattern in Fig.~\ref{fig_FFRpattern}\\
\(f_1=f_2\)\\
Overshoot \(\leq 35\%\)\\
\(\displaystyle
t_{\text{act}}\leq
\begin{cases}
1.3\,\text{s}, & \text{if } f_1=49.7\,\text{Hz},\\
1.0\,\text{s}, & \text{if } f_1=49.6\,\text{Hz},\\
0.7\,\text{s}, & \text{if } f_1=49.5\,\text{Hz}
\end{cases}
\)\\
\(t_{\text{dur}}=5\,\text{s}\ \text{or}\ 30\,\text{s}\)\\
\(\displaystyle
\text{Deactivation: }
\begin{cases}
\text{rate}\leq 20\%/\text{s},
& \text{if } t_{\text{dur}}=5\,\text{s},\\
\text{time}\leq 30\,\text{s},
& \text{if } t_{\text{dur}}=30\,\text{s}
\end{cases}
\)\\
\(P_{\text{res}}\leq 25\%\,\Delta P\)\\
\(t_{\text{res}}\leq 15\,\text{min}\)
} 
\\
        \hline
        \textbf{Procurement} & \makecell{Long-term Y-4: Years-ahead auction\\Mid-term Y-1: Year-ahead auction\\Short-term D-1: Day-ahead auction\\bid $\geq$ 100 $\text{MVA}\cdot\text{s}$} & \makecell{Years ahead bid\\of a fraction \\of the certified inertia\\ for a 2 to 10-year period} & \makecell{Day-ahead auction \\ Period unit: 1 h}  
        \\
        \hline
        \textbf{Remuneration} & \makecell{Availability \& Utilization\\(pay-as-bid dual price basis)} & \makecell{Fixed Price \\ based on availability \\during 15-min intervals }& \makecell{Availability \\ (Marginal Price (Pay-as-clear)\\Regulated Price (only one bid))} \\
        \hline
                \textbf{Stacking} & \makecell{Stackable} & \makecell{Stackable}& \makecell{Stackable} \\
        \hline
        \textbf{Reference} & \cite{NESO_Stability_25_26_ITT_V5,NESO_Stability_26_27_ITT_V2,NESO_Stability_27_28_EOI_V1,NESO_LT2029_ITT_2025} &  \cite{Netztransparenz2026MarktdesignMomentanreserve,Netztransparenz2026_Verfuegbarkeitsbestimmung,Netztransparenz2026_DatenaustauschAbrechnung,Netztransparenz2025_AngebotsabgabeAggregation} & \cite{noauthor_ancillary_2025_energinet} \\
        \hline
    \end{tabular}
  }
    \caption{Commercial service products of inertia and fast frequency reserve}
    \label{tab:IR&FFR_Commc}
\end{table}

\begin{figure*}[!t]
\centering
\subfloat[]{\includegraphics[width=\textwidth]{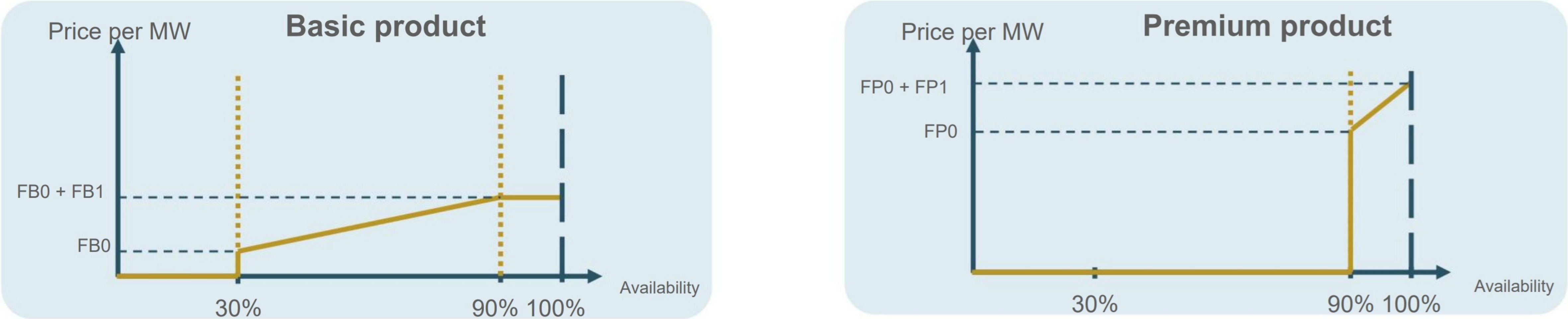}
\label{fig_curves}}
\hfil
\subfloat[]{\includegraphics[width=\textwidth]{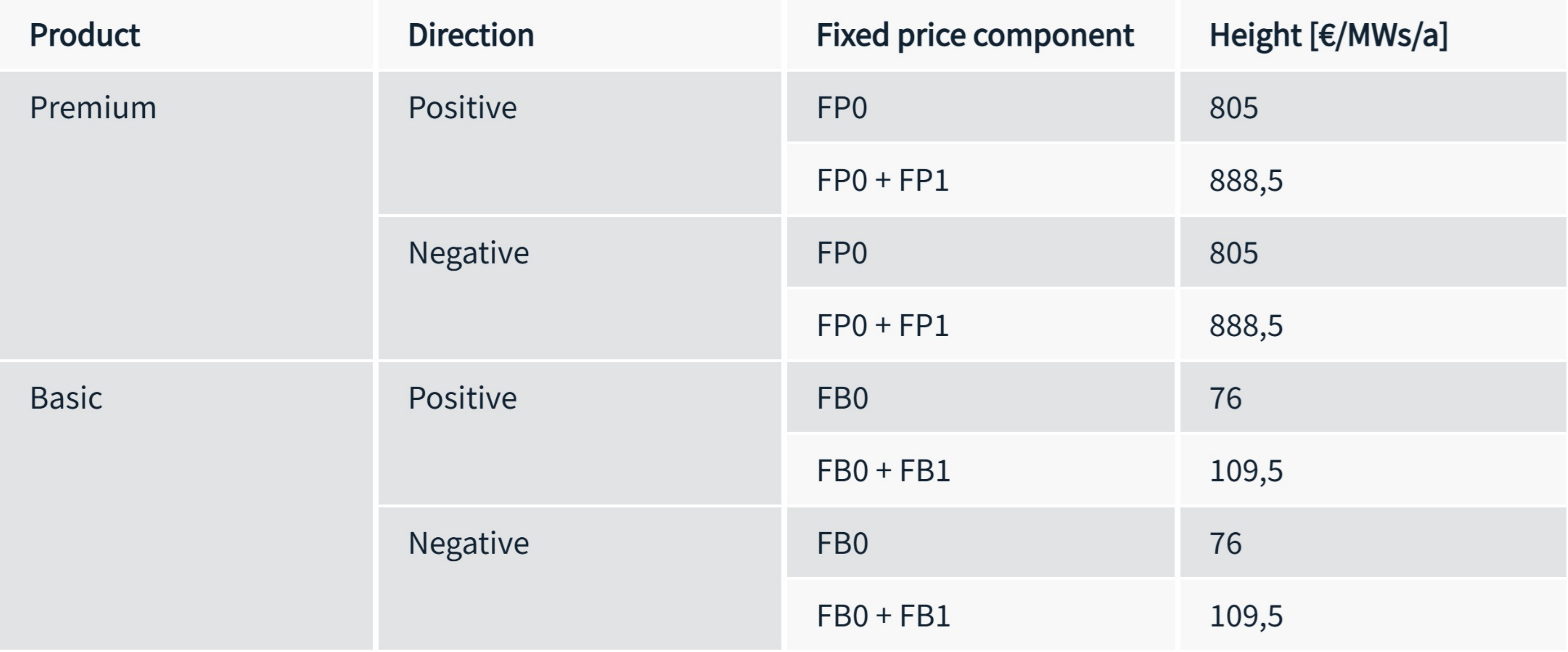}
\label{fig_prices}}
\caption{Remuneration mechanism of the instantaneous reserve (Momentanreserve) market in Germany \cite{netztransparenz_inertia_2026}. (a)curves; (b)prices.}
\label{fig_IR_remuneration_DE}
\end{figure*}

\subparagraph{GB}
The emerging stability market in Great Britain is open to a broad range of potential participants. Eligible units must be directly connected to the transmission system ($\geq132\text{kV}$), meet high availability ($\geq90\%$) requirements, and comply with the full GFM capability requirements specified in the Grid Code. These requirements extend beyond RoCoF response to include fast fault current injection, phase-jump response, active damping power, etc \cite[ECC.6.3.19]{NESO2026_GridCode}. In addition, participating units are required to maintain appropriate reactive power capability commensurate with their unit type and rated capacity. 

The market is highly structured. There are three procurement horizons: short-term (D-1), mid-term (Y-1), and long-term (Y-4). At present, three mid-term tender rounds are being conducted sequentially, the first long-term tender has been initiated, and the short-term market remains under development. Service provision is mandatory in both directions, enabling both upward and downward power response. Remuneration is based on both availability and utilization under a pay-as-bid dual-pricing mechanism.

From a technical perspective, the specifications defined for the stability market remain non-exhaustive. Quantification of the service is primarily based on the inertia time constant and RoCoF of 1 Hz/s, while the dynamic requirement for Active Inertia Power is limited to an initial response delay of no more than 5 ms~\cite{neso_grid_forming_guidance_2026}. Compliance test procedures are outlined in \cite[ECP.A.9]{NESO2026_GridCode}. However, the detailed acceptance criteria applied during assessment are not publicly available.

In the first and second tender rounds of mid-term Y-1 market, contracts were awarded exclusively to synchronous GFM units \cite{NESO_Stability_Y1_Round1_Results_2024}. Non-synchronous units were unsuccessful, as they did not pass the technical question assessment, feasibility report assessment, and eligibility criteria assessment.

\subparagraph{DE}

To enter the instantaneous reserve (Momentanreserve) market in Germany, eligible HVDC-OWPP participants must obtain a GFM certificate corresponding to a specified inertia time constant for each generator or storage unit within the plant. The certification process follows the FNN GFM guidelines \cite{VDEFNN2020GFM,VDEFNN2025GFM}, which are consistent with the technical requirements in grid codes. In actual bidding, only a fraction of the certified inertia is allowed to be offered to the market.

In contrast to the mandatory bi-directional requirement in Great Britain, service provision in Germany is direction-specific and optional, allowing providers to participate either in positive (upward response for under-frequency) or negative (downward response for over-frequency) inertia products (However, Grid-forming units offering both positive and negative inertia shall be designed to be symmetric \cite[4.2.1.12]{VDEFNN2025GFM}). Accordingly, the procured market volume is defined separately for the two directions, with a larger volume currently allocated to negative inertia. This allocation is determined based on characteristic grid-separation scenarios \cite{GermanTSO2025SystemStability}. In addition, the required market capacity is expected to increase in the future, driven by growing large-scale power transits in the AC grid associated with increasingly remote generation. For each direction, there are basic and premium products, where the premium product features stricter availability requirements and correspondingly higher remuneration. Procurement is organized as a years-ahead auction in which providers bid a fraction of their certified inertia for contract durations ranging from 2 to 10 years. 

Availability is assessed based on the 15-minute average active power to verify that sufficient reserve power is maintained within each settlement interval, assured by quality control using minute-resolution measurements. For unit network such as HVDC-OWPPs, availability is assessed on the basis of the aggregated inertia of all the available units. Stacking with other services is permitted, provided that the availability criteria remain satisfied. Remuneration follows a fixed-price mechanism based exclusively on annual availability, independent of actual activation, as illustrated in Fig.~\ref{fig_IR_remuneration_DE}, where the price list is updated every two years.

\subparagraph{DK}
In Denmark, the marketization of GFM technology remains at an early transitional stage \cite{Energinet2026EvidenceStabilityMarket}. Within the GFM concept, inertia is only one of several relevant contributions, alongside system-strength-related functions such as damping and short-circuit current provision, fast fault-current injection, and dynamic and steady-state voltage regulation. Although the technical effectiveness of GFM technology has been validated, the Danish framework is currently evolving from mandatory technical requirements and compliance methodologies toward potential location-sensitive, long-term stability-market mechanisms for capabilities beyond baseline GFM behavior.

This transition has introduced new elements into the formulation of technical requirements for the market. For example, regarding compliance-test arrangements, the Danish approach differs from the ENTSO-E and German frameworks, where the test setup is primarily defined in the time domain. In the Danish framework, the model test bench is specified in both the time and frequency domains \cite{Kwon2026DeploymentGFM}. Consequently, in addition to time-domain assessments, such as envelope-based performance evaluation, frequency-domain single-input single-output (SISO) and multiple-input multiple-output (MIMO) analyses are considered to identify characteristic GFM features and to provide further insight into damping and stability contributions.

Nevertheless, several challenges remain before GFM capabilities can be translated into market requirements \cite{Energinet2026EvidenceStabilityMarket}. These include defining the regulatory baseline and corresponding market incentives, as well as addressing quantification and qualification challenges associated with GFM capabilities. Such challenges arise from measurement sensitivity, nonlinear behavior under current limitation, operating-point-dependent availability, locational effectiveness, and coupling among different capability dimensions.

\subsubsection{FFR}
In 2020, the Nordic transmission system operators, i.e., Energinet, Fingrid, Statnett, and Svenska kraftnät, jointly developed FFR product to address low-inertia conditions. Since its introduction, FFR has been procured mainly through national capacity markets operated by the Nordic TSOs, with activation requirements and procurement volumes coordinated at the Nordic level. The market is typically activated during periods when system inertia is expected to be low, ensuring that sufficient ultra-fast reserves are available to maintain frequency stability. An overview of the FFR market in DK2 area (east part) of Denmark is
summarized in Table~\ref{tab:IR&FFR_Commc}.

\subparagraph{DK}

Following successful prequalification in accordance with the requirements detailed in \cite{energinet_prequalification_2025}, the participant becomes eligible to submit bids to the FFR market. The market size is determined based on potential large frequency deviations that could arise from contingencies such as the outage of major generation units or transmission lines. However, the actual FFR demand varies over time, as it is inversely proportional to the variable level of system inertia. Technically, providers may choose between three predefined activation thresholds, i.e., 49.7 Hz, 49.6 Hz, or 49.5 Hz, offering flexibility in the design of the response strategy. Furthermore, aggregation of multiple units is permitted, and no restrictions are imposed on stackability, as long as sufficient reserve capacity is available.

\begin{figure*}[!t]
\centering
\subfloat[]
{\includegraphics[width=0.5\columnwidth]{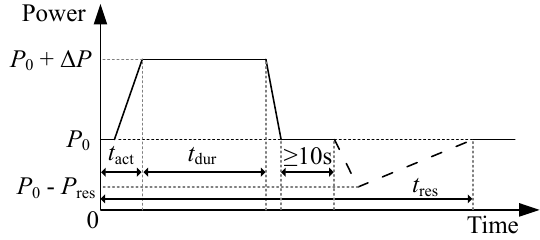}%
\label{fig:FFR_time}}
\hfil
\subfloat[]
{\includegraphics[width=0.3\columnwidth]{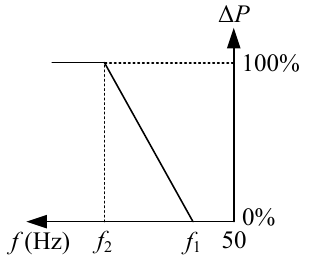}%
\label{fig:FFR_freq}}
    \caption{Response pattern of fast frequency reserve. (a) Time domain. (b) Frequency domain.}
\label{fig_FFRpattern}
\end{figure*}

\section{Frequency Containment Reserve} \label{sec:FCR}


FCR is a fast and short-enduring active power reserve released proportionally to frequency deviation and used to automatically stabilize grid frequency immediately after a disturbance. FCR slows and limits frequency deviations but does not restore the frequency to nominal value. FCR only responses temporarily until FRR restores the frequency deviation to zero. 

\subsection{Grid-code Obligations for Frequency Containment Reserve}
The grid-code requirements for FCR typically encompass a set of frequency control functionalities, namely limited frequency sensitive mode - overfrequency (LFSM-O), limited frequency sensitive mode - underfrequency (LFSM-U), and frequency sensitive mode (FSM). HVDC-OWPPs, as large-scale generation units, are generally required to support all these functionalities with mandatory capabilities, while the actual activation is subject to TSO instructions.

\begin{figure}[htbp]
    \centering
    \includegraphics[width=0.6\textwidth]{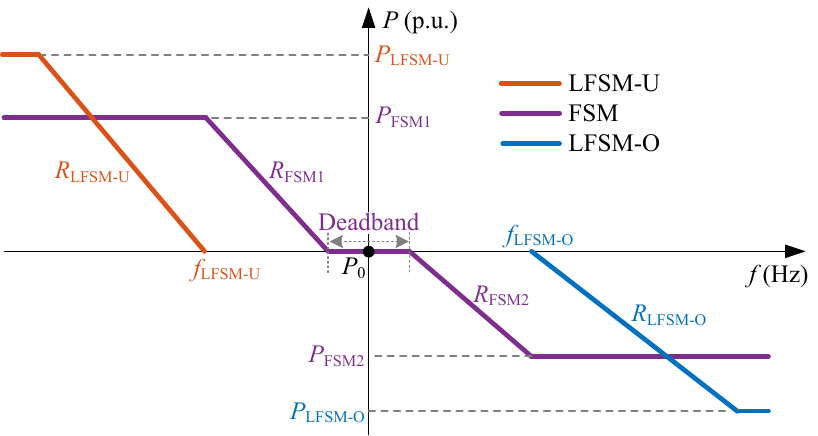} 
    \caption{Frequency-domain response pattern of mandatory frequency containment reserve.}
    \label{fig:FCR_freq}
\end{figure}

\begin{figure}[htbp]
    \centering
    \includegraphics[width=0.6\textwidth]{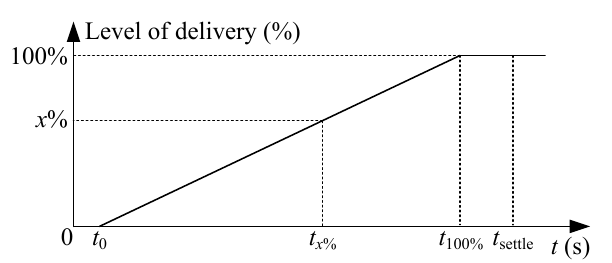} 
    \caption{Time-domain response pattern.}
    \label{fig:FCR_time}
\end{figure}

The grid-code obligations of HVDC-OWPPs for FCR are summarized in Table~\ref{tab:FCR_manda}, Figure~\ref{fig:FCR_freq}, and Figure~\ref{fig:FCR_time}, where Min is the minimum regulating level; Max is the maximum power; $f_{\text{LFSM-O}}$ and $f_{\text{LFSM-U}}$ are the thresholds for LFSM-O and LFSM-U; $P_{\text{LFSM-O}}$, $P_{\text{LFSM-U}}$, $P_{\text{FSM1}}$, and $P_{\text{FSM2}}$ are the power limits for LFSM-O, LFSM-U, and FSM, respectively; $R_{\text{LFSM-O}}$, $R_{\text{LFSM-U}}$, $R_{\text{FSM1}}$, and $R_{\text{FSM2}}$ are the droops for LFSM-O, LFSM-U, and FSM, respectively; $t_{\text{0}}$ is the initial delay; $t_{\text{x\%}}$ is the time for x\% delivery; $t_{\text{settle}}$ is the settling time; P, S, and H are respectively the primary response (P), secondary response
(S), and high frequency response (H) of FSM operation in GB.

\begin{table}[htbp]
    \centering
    \renewcommand{\arraystretch}{1}
    \setlength{\tabcolsep}{6pt}

    \resizebox{\textwidth}{!}{
    \begin{tabular}{|c|*{6}{c|}}
        \hline
        \multicolumn{1}{|c|}{\textbf{}} & \textbf{Service} & \textbf{Target} & \makecell{\textbf{Response Pattern} \\ (based on Figure \ref{fig:FCR_freq} and \ref{fig:FCR_time})} & \textbf{Requirement} & \textbf{Remuneration} & \textbf{Reference} \\
        \hline  

        \multirow{3}{*}{\textbf{\makecell{ENTSO-E \\ (EU)}}} & LFSM-O & All & \makecell{
        $f_{\text{LFSM-O}} \in [50.2, 50.5]\,\text{Hz}$ \\
$R_{\text{LFSM-O}} \in [2, 12]\%$ \\
$P_{\text{LFSM-O}} = \text{Min}$ \\
$t_{0} \leq 2$ s (NC 1.0)\\$t_{95\%/98\%} \leq 2$ s for $50\% P_{\text{max}}$ (IGD for NC 1.0)\\$t_{\text{settle}} \leq 20$ s (IGD for NC 1.0)\\ $t_{90\%} \leq 2$ s for $50\% P_{\text{max}}$  (NC 2.0)
        } & \makecell{Mandatory\\Capability} & \makecell{Not\\Mentioned} & \makecell{
        NC HVDC \cite{noauthor_commission_2016_hvdc}:\\ Art. 39(4)\\
        NC RfG \cite{noauthor_commission_2016_rfg}:\\ Art. 13(2)\\
        IGD for NC 1.0 \cite{ENTSOE2018_LFSM}\\
        NC HVDC 2.0 \cite{ACER2024HVDCAmendment}:\\ Art. 39(4)\\
        NC RfG 2.0 \cite{ACER2023RfGAmendment}:\\ Art. 13(3)}\\

        \cline{2-7}
        
        & LFSM-U & \makecell{Type \\ C \& D} & \makecell{
        $f_{\text{LFSM-U}} \in [49.5, 49.8]\,\text{Hz}$ \\
        $R_{\text{LFSM-U}} \in [2, 12]$\% \\
        $P_{\text{LFSM-U}} = \text{Max}$ \\
        $t_{0} \leq 2$ s (NC 1.0)\\
        $t_{95\%/98\%} \leq 5$ s
        for $20\% P_{\text{max}}$ and $P_{\text{0}} \geq 50\%$ \\(IGD for NC 1.0)\\
        $t_{\text{settle}} \leq 30$ s (IGD for NC 1.0)\\
        $t_{90\%} \leq 10$ s for $50\% P_{\text{max}}$  (NC 2.0)
        } & \makecell{Mandatory\\Capability} & \makecell{Not\\Mentioned} &  \makecell{
        NC HVDC \cite{noauthor_commission_2016_hvdc}:\\ Art. 39(7)\\
        NC RfG \cite{noauthor_commission_2016_rfg}:\\ Art. 15(2)(c)\\
        IGD for NC 1.0 \cite{ENTSOE2018_LFSM}\\
        NC HVDC 2.0 \cite{ACER2024HVDCAmendment}:\\ Art. 39(7)\\
        NC RfG 2.0 \cite{ACER2023RfGAmendment}:\\ Art. 15(2)(c)} \\
        
        \cline{2-7}
        
        & FSM & \makecell{Type \\ C \& D} & \makecell{
        $R_{\text{FSM1}} = R_{\text{FSM2}} \in [2, 12]$\% (NC 1.0)\\
        $R_{\text{FSM1}} = R_{\text{FSM2}} \in [2, 27]$\% (NC 2.0)\\
        Insensitivity (NC 1.0): $\pm [10, 30]\,\text{mHz}$ \\
        Deadband (NC 1.0): [0, 500] mHz\\
        ( Insensitivity + Deadband ) (NC 2.0): \\$\pm[10,15]\,\text{mHz}$ \\
        $P_{\text{FSM1}} = \min\{P_r+\delta P_r, \text{Max}\}$, $\delta \in [1.5, 10]$\% \\
        $P_{\text{FSM2}} = \max\{P_r-\delta P_r, \text{Min}\}$, $\delta \in [1.5, 10]$\% \\
        $t_{0} \leq 0.5$ s; $t_{100\%} \leq 10...30$ s (NC 1.0 \& IGD) \\$t_{100\%} \leq 5$ s for IE \& NI (NC 2.0) 
        } & \makecell{Mandatory\\Capability} & \makecell{Not\\Mentioned} &  \makecell{
        NC HVDC \cite{noauthor_commission_2016_hvdc}:\\ Art. 39(8)\\
        NC RfG \cite{noauthor_commission_2016_rfg}:\\ Art. 15(2)(d)\\
        IGD for NC 1.0 \cite{ENTSOE2018_FSM}\\
        NC HVDC 2.0 \cite{ACER2024HVDCAmendment}:\\ Art. 39(9)\\
        NC RfG 2.0 \cite{ACER2023RfGAmendment}:\\ Art. 15(2)(d)} \\
        \hline

       \multirow{3}{*}{\textbf{\makecell{NESO \\ (GB)}}} & LFSM-O & All 
       & \makecell{$f_{\text{LFSM-0}} = 50.4\,\text{Hz}$\\
$R_{\text{LFSM-0}} \in [2, 10]\%$ \\
$P_{\text{LFSM-0}} = \text{Min}$ \\
$t_{0} \leq 2$ s; $t_{100\%} \leq 10$ s ... 3 min} 
       & \makecell{Always On\\ (exc. FSM)} & None & \makecell{Grid Code \cite{NESO2026_GridCode}:\\ECC.6.3.7.1\\
       Grid Code \cite{NESO2026_GridCode}:\\BC3.7.2.2
       }\\
       
        \cline{2-7}
        
        & LFSM-U & \makecell{Type \\ C \& D} 
        &\makecell{$f_{\text{LFSM-U}} = 49.5\,\text{Hz}$ \\
$R_{\text{LFSM-U}} \leq 10\%$ \\
$P_{\text{LFSM-U}} = \text{Max}$ \\
$t_{0} \leq 2$ s} 
        & \makecell{Provision if\\ Inherent\\Capability\\Available\\(exc. FSM)} & None &
        \makecell{Grid Code \cite{NESO2026_GridCode}:\\ECC.6.3.7.2\\
       Grid Code \cite{NESO2026_GridCode}:\\BC3.7.2.2
       }
       \\
        
        \cline{2-7}
        
        & FSM & \makecell{Type \\ C \& D} 
        & \makecell{
$R_{\text{FSM1}}=R_{\text{FSM2}}\in[3,5]\%$\\
$\text{Insensitivity}=\pm0.03\%\times50\,\text{Hz}$\\
$\text{Deadband}=0\,\text{mHz}$\\
$P_{\text{FSM1}}=\min\{P_0+10\%P_n,\text{Max}\}$\\
$P_{\text{FSM2}}=\max\{P_0-10\%P_n,\text{Min}\}$\\
$t_0\leq1\,\text{s};
\quad t_{100\%}\leq10\,\text{s}\ \text{(P \& H)},\
30\,\text{s}\ \text{(S)}$\\
Duration: 30 s (P); 3 min (S)
} 
        & \makecell{Mandatory\\Provision\\Activated by\\Instruction} & \makecell{Holding\\Payment\\ \& \\Response\\Energy\\Payments} &
        \makecell{Grid Code \cite{NESO2026_GridCode}:\\ECC.6.3.7.3\\
       Grid Code \cite{NESO2026_GridCode}:\\BC3.5.4\\
       CUSC \cite{ESO_2025_CUSC}:\\4.1.3.8
       }\\
        \hline

        \multirow{3}{*}{\textbf{\makecell{Tennet \\ (DE)}}} & LFSM-O & All 
        & \makecell{$f_{\text{LFSM-0}} = 50.2\,\text{Hz}$ \\
$R_{\text{LFSM-0}} = 5\%$;
$P_{\text{LFSM-0}} = \text{Min}$ \\
$t_{90\%} \leq 2$ s ($\Delta P \leq 50\%P_{\text{n}}$)\\
$t_{0} \leq 2$ s; $t_{\text{settle}} \leq 20$ s} 
        & \makecell{Always On} & \makecell{Not\\Mentioned} &
        \makecell{NAR \cite{NAR}:\\10.2.3.4.5 \\
        TCR HVDC \cite{vde4131}:\\10.2.5.3 } \\
        
        \cline{2-7}
        
        & LFSM-U & All 
        & \makecell{$f_{\text{LFSM-U}} = 49.8\,\text{Hz}$ \\
$R_{\text{LFSM-U}} = 5\%$;
$P_{\text{LFSM-U}} = \text{Max}$ \\
$t_{90\%} \leq 5\,\text{s} \quad (\Delta P \leq 20\%P_n)$ \\
$t_{90\%} \leq 10\,\text{s} \quad (\Delta P \leq 50\%P_n)$ \\
$t_{0} \leq 2$ s; $t_{\text{settle}} \leq 30\,\text{s}$} 
        & \makecell{Always On} & \makecell{Not\\Mentioned} & \makecell{NAR \cite{NAR}:\\10.2.3.4.5 \\
        TCR HVDC \cite{vde4131}:\\10.2.5.4 } \\
        
        \cline{2-7}
        
        & FSM & All 
        & \makecell{
        $R_{\text{FSM1}} = R_{\text{FSM2}}
= \SI{6}{\percent}
\in [2,12]\%$
\\
$\text{Deadband}
= \SI{200}{\milli\hertz}
\leq \pm \SI{0.04}{\percent}
\times \SI{50}{\hertz}$
\\
$P_{\text{FSM1}}
= \min\{P_0+\delta P_{\text{avai}},\text{Max}\}$
\\
$P_{\text{FSM2}}
= \max\{P_0-\delta P_{\text{avai}},\text{Min}\}$
\\
$\delta=2\% \in [1.5,10]\%$
\\
$t_0\leq2\,\text{s};
\quad t_{100\%}\leq30\,\text{s}$} 
        & \makecell{Always On} & \makecell{Not\\Mentioned} & \makecell{NAR \cite{NAR}:\\10.2.3.4.5 \\
        TCR HVDC \cite{vde4131}:\\10.2.5.2 } \\
        \hline

 \multirow{3}{*}{\textbf{\makecell{Energinet \\ (DK)}}} & LFSM-O & All & \makecell{
        $f_{\text{LFSM-O}} = 50.2\text{ Hz (DK1)}; 50.5\text{ Hz (DK2)}$ \\
$R_{\text{LFSM-O}} = 5\%\text{ (DK1)}; 4\%\text{ (DK2)}$ \\
$P_{\text{LFSM-O}} = \text{Min}$ \\
$t_{0} \leq 2$ s
        } & \makecell{Mandatory\\Capability} & \makecell{Not\\Mentioned} & \makecell{
        NC HVDC (DK)\\ \cite{Energinet2019_HVDC_Requirements}: Art. 39(4)\\
        NC RfG (DK)\\ \cite{Energinet2025_NC_RfG_V5}: Art. 13(2)}\\

        \cline{2-7}
        
        & LFSM-U & \makecell{Type \\ C \& D} & \makecell{
        $f_{\text{LFSM-U}} = 49.8\text{ Hz (DK1)}; 49.5\text{ Hz (DK2)}$ \\
$R_{\text{LFSM-U}} = 5\%\text{ (DK1)}; 4\%\text{ (DK2)}$ \\
$P_{\text{LFSM-U}} = \text{Max}$ \\
$t_{0} \leq 2$ s} & \makecell{Mandatory\\Capability} & \makecell{Not\\Mentioned} &  \makecell{
        NC HVDC (DK)\\ \cite{Energinet2019_HVDC_Requirements}: Art. 39(7)\\
        NC RfG (DK)\\ \cite{Energinet2025_NC_RfG_V5}: Art. 15(2)(c)} \\
        
        \cline{2-7}
        
        & FSM & \makecell{Type \\ C \& D} & \makecell{
        $R_{\text{FSM1}} = R_{\text{FSM2}} \in [2, 12]$\% \\
        Insensitivity = 10 mHZ \\
        Deadband $\in$ [0, 200] mHz (DK1)\\Deadband $\in$ [0, 500] mHz (DK2)\\
        $P_{\text{FSM1}} = \min\{P_r+\delta P_r, \text{Max}\}$, $\delta \in [1.5, 10]$\% \\
        $P_{\text{FSM2}} = \max\{P_r-\delta P_r, \text{Min}\}$, $\delta \in [1.5, 10]$\% \\
        $t_{0} \leq 2$ s; $t_{100\%} \leq 30$ s} & \makecell{Mandatory\\Capability} & \makecell{Not\\Mentioned} &  \makecell{
        NC HVDC (DK) \\ \cite{Energinet2019_HVDC_Requirements}: Art. 39(8)\\
        NC RfG (DK)\\ \cite{Energinet2025_NC_RfG_V5}: Art. 15(2)(d)} \\
        \hline

    \end{tabular}
    }
    \caption{Grid-code obligations of HVDC-OWPPs for frequency containment reserve services}
    \label{tab:FCR_manda}
\end{table}

\subparagraph{EU}
At the EU level, the transition from NC 1.0 to the forthcoming NC 2.0 preserves the required frequency response range, while tightening the requirements on response speed. Specifically, the activation thresholds $f_{\text{LFSM-O}}$ and $f_{\text{LFSM-U}}$ remain unchanged, defining the frequency deviations at which LFSM-O and LFSM-U are initiated until absolute power limits are reached. Similarly, $P_{\text{FSM1}}$ and $P_{\text{FSM2}}$ continue to impose symmetric caps of up to $\pm10\%$ on FSM provision. Notably, the allowable droop range for FSM is expected to increase, despite both the frequency activation range and the provision capacity limits remaining unchanged. On the other hand, the stringent response time requirements for LFSM-O and LFSM-U specified in the IGD under NC 1.0 are expected to be adapted and formalized in NC 2.0. This is accompanied by updated response speed requirements for FSM in low-inertia systems, such as Ireland (IE) and Northern Ireland (NI). 

Furthermore, for FSM, the present requirement on insensitivity alone is anticipated to be replaced by a constraint on the maximum combined effect of inherent frequency response insensitivity and any intentional frequency response deadband, with an even more stringent upper bound. This revision is motivated by the fact that an excessive combined effect can distort the probability density of system frequency and undermine the objective of achieving equal FCR provision over frequency deviations \cite{ENTSOE2018_FSM}.

\subparagraph{DE}
In Germany, under Tennet's regulation, the requirements on provision capacity and response speed for LFSM and FSM are generally aligned with those specified in NC 1.0, with LFSM response speed adhering to the more stringent criteria defined in the IGD under NC 1.0. Continuous operation in both LFSM and FSM is required. Due to the deadzone setting of FSM, the activation ranges of FSM and LFSM may overlap in practice. While the FSM provision necessitates curtailment to reserve upward response capacity, however, the mandated provision capacity is limited to 2\%, representing a relatively minor requirement. Without mentioned explicitly, no remuneration is available, as these services are mandatory ancillary services.

\subparagraph{GB}
\label{FCR_Manda_GB}
In Great Britain, the requirements on provision capacity and response speed for LFSM and FSM are generally aligned with those specified in NC 1.0, while LFSM and FSM are mutually exclusive in operation, and HVDC-OWPP should operate in either LFSM or FSM, with LFSM as the default mode and FSM activated upon instruction, which may be issued at any time. 

Furthermore, the response under FSM operation is categorized into three components: primary response (P), secondary response (S), and high-frequency response (H). The primary and secondary responses address under-frequency events and follow the corresponding frequency-domain response pattern in the under-frequency range, whereas the high-frequency response applies to over-frequency events and follows the corresponding response pattern in the over-frequency range. The primary response provides a fast, short-duration contribution, while the secondary response is slower but sustained over a longer period. The required response components are specified in the FSM instruction.

The energy utilization of LFSM and FSM follows the principle of either relying on inherent capability or being subject to remuneration. For LFSM-O, HVDC-OWPPs, as generation units, inherently possess footroom when operating above their minimum regulation level. Therefore, this inherent capability is utilized without additional compensation by NESO. For LFSM-U, provision is required only when inherent capability is available (e.g., when the HVDC-OWPP is already curtailed due to excessive wind power or other ancillary service commitments such as FRR, RR, etc.), and thus no dedicated curtailment is imposed. In contrast, FSM provision is symmetric and requires HVDC-OWPP to reserve headroom for the upward response, which necessitates active curtailment. The associated energy loss is compensated through holding payments covering the entire duration of FSM activation, determined on a pay-as-bid basis. In addition, the actual response delivered during FSM is remunerated via energy-based payments. 

As mentioned above, a salient feature of FSM provision in GB is its integration with market-based mechanisms. Generators submit their response capability and associated holding price on a monthly basis, with a forecast horizon of approximately 15 to 45 days. By default, the reference price for energy response payment is zero for HVDC-OWPPs, as they are classified as non-fuel-cost power stations. However, an exception applies to units registered as “CfD BMUs,” which may make a one-time election at the start of their agreement to set the reference price for response energy payments based on market index price for the duration of that agreement \cite[4.1.3]{ESO_2025_CUSC}.

\subparagraph{DK}
Although no HVDC-OWPP or any other HVDC-connected power park modules are currently in operation in Denmark, NC 1.0 has already been adopted in Denmark in anticipation of future developments, such as the Bornholm energy island and the North Sea Energy Island. The Danish adapted version \cite{Energinet2019_HVDC_Requirements,Energinet2025_NC_RfG_V5} follows NC 1.0, with more specific parameter settings for the two synchronous areas: DK1 (connected to continental Europe) and DK 2 (connected to the Nordic system), as well as more detailed requirements on control setting resolution and measurement accuracy.

While the Danish adaptation of NC 1.0 primarily specifies mandatory capabilities, potential implementation approaches can be inferred from the requirements applied to existing wind power plants. According to TR 3.2.5 \cite{energinet_Guidelines_TR325_2016} which applies to wind power plants above 11 kW in Denmark, LFSM operation is limited to LFSM-O (i.e., Frequency Response in TR 3.2.5), which must be continuously active, with requirements aligned with NC 1.0. In contrast, FSM (i.e., Frequency Control in TR 3.2.5) activation is subject to agreement with Energinet, and exhibits two main deviations from NC 1.0. First, distinct frequency response characteristics are defined in the overfrequency range to provide required flexibility. Second, the upward provision capacity may reach up to 20\% of nominal capacity, which is twice the corresponding limit in NC 1.0.

\subsection{Commercial Products of Frequency Containment Reserve}
\begin{figure}[htbp]
    \centering
    \includegraphics{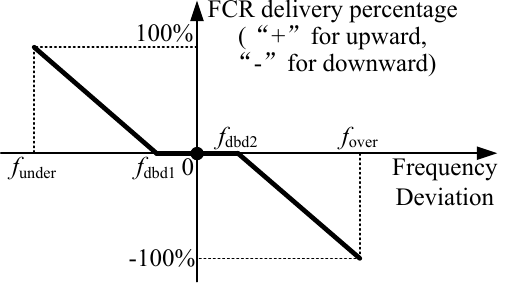} 
    \caption{Generic frequency-domain response pattern of commercial frequency containment reserve service.}
    \label{fig:FCR_Comm}
\end{figure}

Although European-level regulations have established a general framework for technical guidelines \cite{EU2017_2195} and market mechanisms \cite{EU2019_943}, governing balancing services including FCR, variations in implementation and exemptions persist across different countries and districts. These differences warrant a case-by-case analysis. Accordingly, the commercial FCR services in Great Britain, Germany, and Denmark are summarized in Table~\ref{tab:FCR_Commercial}.
In addition, a generic frequency-domain response pattern of commercial FCR is illustrated in Figure~\ref{fig:FCR_Comm}, where $f_{\text{dbd}1}$ and $f_{\text{dbd}1}$ are the frequency deadband boundary; $P_{\text{dbd}1}$ and $P_{\text{dbd}1}$ are the power deadband boundary; $f_{\text{up}}$ and $f_{\text{dwn}}$ are the delivery boundary on frequency. In general, FCR services on all platforms are procured through day-ahead auctions and primarily remunerated based on availability under a pay-as-clear mechanism.

\begin{table}[htbp]
    \centering
    \renewcommand{\arraystretch}{1}
    \setlength{\tabcolsep}{6pt}

    \resizebox{\textwidth}{!}{
    \begin{tabular}{|c|*{8}{c|}}
        \hline
        \multicolumn{1}{|c|}{\textbf{}} & \textbf{Service} & \textbf{Direction} & \textbf{Market} & \textbf{\makecell{Technical\\Requirements}} & \textbf{Procurement} & \textbf{Remuneration}& \textbf{Stacking} & \textbf{Reference} \\
        \hline  

        \multirow{3}{*}{\textbf{\makecell{NESO\\ (GB)}}} 
        & \makecell{Dynamic\\ Containment\\ (DC)} 
        & \makecell{Asym-\\metric}
        & \makecell{Bid Size:\\$[1, 100]$ MW} 
        & \makecell{ Figure \ref{fig:FCR_Comm_GB} and \ref{fig:FCR_time}\\ $t_{0} = 0.5$ s\\ $t_{100\%} = 1$ s} 
        & \makecell{Auction:\\D-1;\\H-1 (developing)\\Period Unit: 4h;\\30 min (developing)}
        & \makecell{Availability\\D-1: Pay-as-clear;\\H-1: Pay-as-bid\\(developing)}
        & Stackable
        & \makecell{\cite{ESO_DCDMDR_2024}  \cite{ESO_FCR_procurement_2024}\\
\cite{ESO_FCR_service_term_2024} 
        \cite{NESO_DynamicResponse_RealTime}}\\

        \cline{2-9}
        
        & \makecell{Dynamic\\ Moderation\\ (DM)} 
        & \makecell{Asym-\\metric}
        & \makecell{Bid Size:\\$[1, 100]$ MW} 
        & \makecell{ Figure \ref{fig:FCR_Comm_GB} and \ref{fig:FCR_time} \\ $t_{0} = 0.5$ s\\ $t_{100\%} = 1$ s}
        & \makecell{Day-ahead\\Auction\\Period Unit: 4h;\\30 min (developing)}
        & \makecell{Availability\\D-1: Pay-as-clear;\\H-1: Pay-as-bid\\(developing)}
        & Stackable
        &  \makecell{\cite{ESO_DCDMDR_2024}  \cite{ESO_FCR_procurement_2024}\\
\cite{ESO_FCR_service_term_2024} 
        \cite{NESO_DynamicResponse_RealTime}}\\
        
        \cline{2-9}
        
        & \makecell{Dynamic\\ Regulation\\ (DR)} 
        & \makecell{Asym-\\metric}
        & \makecell{Bid Size:\\$[1, 100]$ MW} 
        & \makecell{ Figure \ref{fig:FCR_Comm_GB} and \ref{fig:FCR_time} \\ $t_{0} = 2$ s\\ $t_{100\%} = 10$ s}
        & \makecell{Day-ahead\\Auction\\Period Unit: 4h;\\30 min (developing)}
        & \makecell{Availability\\D-1: Pay-as-clear;\\H-1: Pay-as-bid\\(developing)}
        & Stackable
        &  \makecell{\cite{ESO_DCDMDR_2024}  \cite{ESO_FCR_procurement_2024}\\
\cite{ESO_FCR_service_term_2024} 
        \cite{NESO_DynamicResponse_RealTime}}\\

        \cline{2-9}
        
        & \makecell{Static\\ Firm\\Frequency \\ Response\\(SFFR)} 
        & Upward
        & \makecell{Market\\Capacity:\\250 MW\\Bid size (MW):\\$[1, \text{max}]$ (now)\\$[0.1, 100]$ (new)}
        & \makecell{ Figure \ref{fig:SFFR_ESO} and \ref{fig:FCR_time} \\ $t_{100\%} = 30$ s \\ Duration = 30 min\\Trigger Frequency:\\49.7 Hz (now)\\49.65 Hz (new)}
        & \makecell{Day-ahead\\Auction\\Period Unit:\\4h}
        & \makecell{Availability\\(Pay-as-clear)}
        & Stackable
        & \cite{ESO_SFFR_ServiceTerm_2023,ESO_SFFR_procurement_2023,NESO2026_StaticFFR_Submission}\\
        \hline

\textbf{\makecell{Tennet\\ (DE)}}
& \makecell{FCR\\(FCR\\ Cooperation)} 
        & \makecell{Sym-\\metric}
        &\makecell{Market\\Capacity:\\ $\sim$564 MW\\Bid Size:\\$[1, 25]$ MW }
        & \makecell{Figure \ref{fig:FCR_Comm} and \ref{fig:FCR_time}\\$f_{\text{dbd1}} = -10$ mHz\\$f_{\text{dbd2}}
 = 10$ mHz\\$f_{\text{under}} = -200$ mHz\\$f_{\text{over}} = 200$ mHz\\$t_{50\%} = 15$s\\$t_{100\%} = 30$s}
        & \makecell{Day-ahead\\Auction\\Period Unit:\\4h}
        & \makecell{Availability\\(Pay-as-clear)}
        & Stackable
        & \makecell{\cite{noauthor_wwwregelleistungnet_nodate, ENTSOE_FCRcooperation}\\\cite{ENTSOE_TSO2017/1485_2017}:\\Art. 154(1)}\\

        \hline
        
        \multirow{3}{*}{\textbf{\makecell{Energinet\\ (DK)}}} 
        & \makecell{FCR\\(DK1)} 
        & \makecell{Sym-\\metric}
        & \makecell{Market\\Capacity:\\$\pm 25 $MW\\(DK)\\$\pm 100 $MW\\(Export)\\Bid Size:\\$\geq 1$ MW}
        & \makecell{Figure \ref{fig:FCR_Comm} and \ref{fig:FCR_time}\\$f_{\text{dbd1}} \in [-20, 0]$ mHz \\$f_{\text{dbd2}} \in [0, 20]$ mHz\\$f_{\text{under}} = -200$ mHz\\$f_{\text{over}} = 200$ mHz\\$t_{0} = 2$s\\$t_{50\%} = 15$s\\$t_{100\%} = 30$s}
        & \makecell{Day-ahead\\Auction\\Period Unit:\\4 h}
        & \makecell{Availability\\(Pay-as-clear);\\supplied energy\\settled as\\imbalances}
        & Stackable
        & \cite{noauthor_ancillary_2025_energinet,energinet_prequalification_2025}\\

        \cline{2-9}
        
        & \makecell{FCR-N\\(DK2)} 
        & \makecell{Sym-\\metric}
        & \makecell{Market\\Capacity:\\$\pm 19 $MW\\(DK)\\Bid Size:\\$\geq 0.1$ MW}
        & \makecell{Figure \ref{fig:FCR_Comm} and \ref{fig:FCR_time}\\$f_{\text{dbd1}} = f_{\text{dbd2}}
 = 0$\\$f_{\text{under}} = -100$ mHz\\$f_{\text{over}} = 100$ mHz\\$t_{0} = 2.5$s\\$t_{63\%} = 60$s\\$t_{95\%} = 3$min}
        & \makecell{Day-ahead\\Auction;\\Period Unit:\\1 h\\ (up to 3 or 6 h)}
        & \makecell{Availability\\(Pay-as-clear)\\ \& \\Utilisation\\(based on\\mFRR\\average price)}
        & Stackable
        & \cite{noauthor_ancillary_2025_energinet,energinet_prequalification_2025}\\
        
        \cline{2-9}
        
        & \makecell{FCR-D\\(DK2)} 
        & \makecell{Asym-\\metric}
        & \makecell{Market\\Capacity:\\$+ 46 $MW\\$- 44 $MW\\Bid Size:\\$\geq 0.1$ MW}
        & \makecell{Figure \ref{fig:FCR_Comm} and \ref{fig:FCR_time}\\$f_{\text{dbd1}}
 = -100$mHz\\$f_{\text{dbd2}}
 = 100$mHz\\$f_{\text{up}} = -500$ mHz\\$f_{\text{dwn}} = 500$ mHz\\$t_{0} = 2.5$s\\$t_{86\%} = 7.5$s\\$E_{\text{7.5s}} = S_{\text{c}}\cdot3.2$s}
        & \makecell{Day-ahead\\Auction;\\Period Unit:\\1 h\\ (up to 3 or 6 h)}
        & \makecell{Availability\\(Pay-as-clear);\\supplied energy\\settled as\\imbalances}
        & Stackable
        & \cite{noauthor_ancillary_2025_energinet,energinet_prequalification_2025}\\

        \hline

    \end{tabular}
    }
    \caption{Commercial services of frequency containment reserve}
    \label{tab:FCR_Commercial}
\end{table}

\begin{figure}[htbp]
    \centering
    \includegraphics[width=0.6\textwidth]{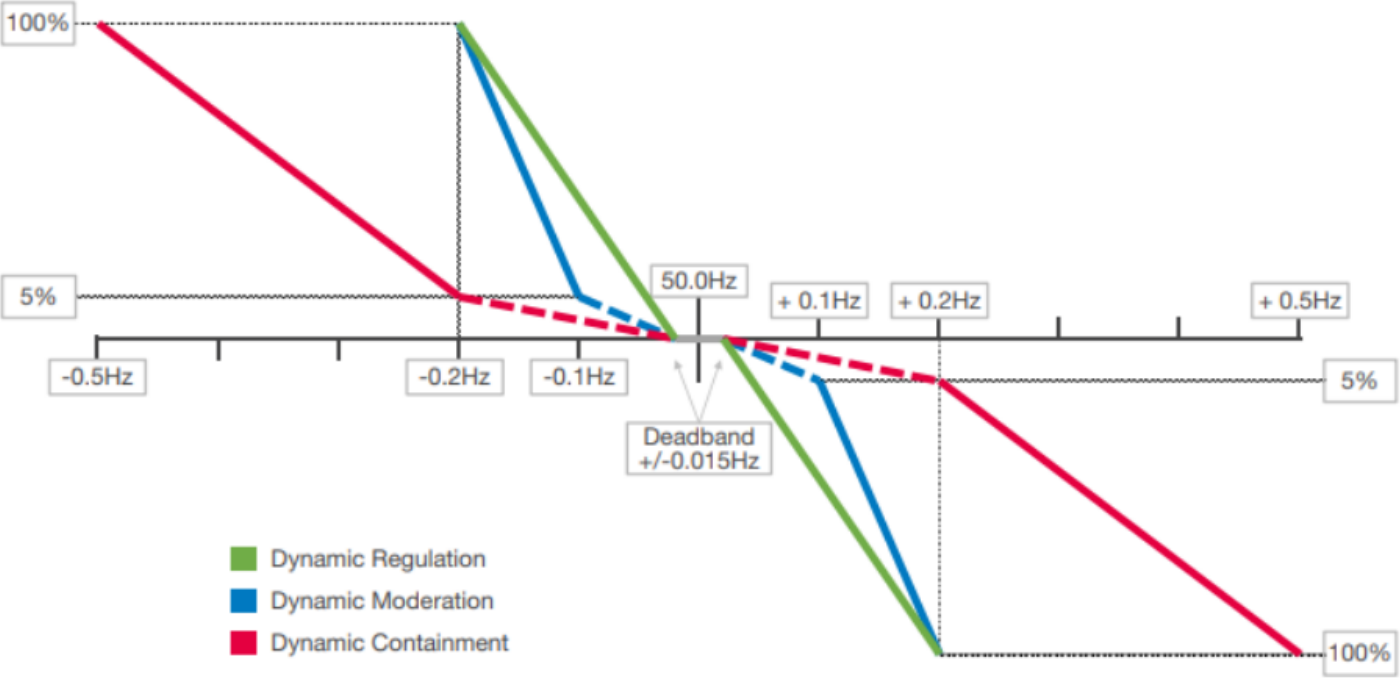} 
    \caption{Frequency-domain response pattern of dynamic services in Great Britain \cite{ESO_DCDMDR_2024}.}
    \label{fig:FCR_Comm_GB}
\end{figure}

\begin{figure}[htbp]
    \centering
    \includegraphics[width=0.6\textwidth]{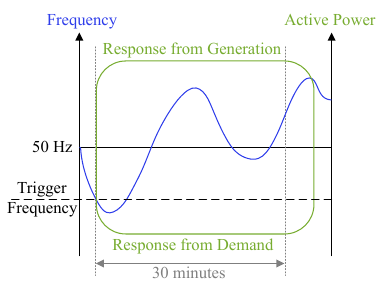} 
    \caption{Response pattern of static firm frequency response required by ESO (GB) \cite{ESO_SFFR_ServiceTerm_2023}.}
    \label{fig:SFFR_ESO}
\end{figure}

\subparagraph{GB}
In Great Britain, a diverse portfolio of commercial FCR-related services is available, including Mandatory and Commercial Frequency Response (MFR/CFR), dynamic frequency services, and Static Firm Frequency Response (SFFR). These services offer flexibility in response direction, speed, and duration, covering different frequency ranges, and are currently undergoing significant reform.

As MFR has been introduced in Section~\ref{FCR_Manda_GB}, providers capable of delivering frequency response beyond their mandatory obligations may enter into bilateral agreements with NESO to provide additional capacity, referred to as Commercial Frequency Response (CFR). The technical requirements and market arrangements for CFR are identical to those for MFR, and thus the relevant specifications are not repeated in Table~\ref{tab:FCR_Commercial}. NESO is currently initiating a comprehensive reform of both MFR and CFR \cite{NESO_ResponseReform_2025_Webinar}. Although the detailed service design phase has not yet commenced, several aspects are under evaluation, including partial activation (i.e., primary, secondary, or high frequency response alone), service duration (open- versus fixed-ended), procurement granularity (e.g., 4-hour versus 30-minute blocks), submission timelines (day-ahead versus hour-ahead), and payment structures.

The dynamic frequency services comprise Dynamic containment (DC), dynamic moderation (DM), and dynamic regulation (DR). DM provides fast-acting pre-fault response during periods of high volatility, while DR represents a slower, steady-state pre-fault service. In contrast, DC is positioned as a post-fault service. Overall, these dynamic services are asymmetric and operate on significantly faster timescales (typically within 1–10 s) compared to conventional commercial FCR services (on the order of around 30 s). Recently, several enhancements to dynamic response services have been proposed to improve operational effectiveness \cite{NESO_DynamicResponse_RealTime}. Key areas under consideration include locational procurement, near real-time response, and shorter procurement intervals. In particular, participants may be able to update their maximum available capacity and corresponding prices closer to real time, i.e., up to one hour before delivery (H-1), while activation instructions would be issued at least two minutes prior to delivery. In addition, procurement block durations are expected to be reduced from 4 hours to 30 minutes. 

SFFR is a non-dynamic (stepwise), post-event service with relatively long delivery duration, designed solely for upward provision. Over the past 18 months, NESO has been developing an update to SFFR to encourage broader participation \cite{NESO2026_StaticFFR_Submission}. The minimum bid size has been reduced from 1 MW to 0.1 MW, and a maximum cap of 100 MW is imposed on a single unit, while the maximum sell size must not exceed the aggregate registered capacity of the associated eligible assets. The response speed and duration requirements remain unchanged, whereas the trigger frequency has been slightly lowered. With respect to under-delivery, availability payments are subject to penalties based on the Percentage Performance Measure (PPM), which is being revised to more accurately reflect service under-performance, as shown in Table~\ref{tab:SFFR_Penalty}. In addition, the updated framework clarifies that the service must still be fully delivered if activation occurs at the end of service block.

\begin{table}[h!]
\centering

\begin{subtable}[t]{0.48\textwidth}
\centering
 \resizebox{\linewidth}{!}{
\begin{tabular}{|c|c|}
\hline
\textbf{\makecell{Percentage\\Performance\\Measure}} & \textbf{\makecell{\% by which \\Availability Payment\\is reduced}} \\ \hline
$\geq 95\%$ & $0\%$ \\ \hline
$\geq 60\%, < 95\%$ & $25\%$ \\ \hline
$\geq 10\%, < 60\%$ & $50\%$ \\ \hline
$< 10\%$ & $100\%$ \\ \hline
\end{tabular}
}
\caption{}
\end{subtable}
\hfill
\begin{subtable}[t]{0.48\textwidth}
\centering
 \resizebox{\linewidth}{!}{
\begin{tabular}{|c|c|}
\hline
\textbf{\makecell{Percentage\\Performance\\Measure}} & \textbf{\makecell{\% by which \\Availability Payment\\is reduced}} \\ \hline
$\geq 95\%$ & $0\%$ \\ \hline
$\geq 75\%, < 95\%$ & \makecell{1\% - 99\%\\(reducing
linearly)} \\ \hline
$< 75\%$ & $100\%$ \\ \hline
\end{tabular}
}
\caption{}
\end{subtable}

\caption{Under-delivery penalty of static firm frequency response in Great Britain. (a) Present; (b) Coming Update.}
\label{tab:SFFR_Penalty}
\end{table}

\subparagraph{DE}

In Germany, commercial FCR is procured through FCR Cooperation \cite{noauthor_wwwregelleistungnet_nodate}, a cross-border market mechanism through which European TSOs jointly procure FCR capacity via a single harmonized auction. Germany contributes a significant share of the total required capacity due to its substantial generation share within the continental European system. The FCR product is symmetric, and its response speed and frequency range requirements are broadly aligned with the FSM requirements specified in NC 1.0. The provision of balancing capacity and the activation (energy delivery) are remunerated through the service fee (availability payment), with no separate compensation for activation, and the associated energy is neutralized in the imbalance settlement.

\subparagraph{DK}
The two synchronous areas in Denmark, DK1 and DK2, are connected to continental Europe and the Nordic system, respectively. Depending on the connection point, HVDC-OWPPs may participate in the corresponding FCR markets. 

In DK1, connected to the continental European synchronous area, commercial FCR is also procured through FCR Cooperation. The FCR product is symmetric, with relatively limited domestic capacity requirements in Denmark determined by its generation share within the continental system, alongside additional capacity potentially available for export. The response speed and frequency range requirements are aligned with the collective specifications, with slightly looser constraint on the combined effect of insensitivity and deadband. While availability is remunerated, the delivered energy is settled as imbalances.
 
In DK2, connected to the Nordic synchronous area, FCR is divided into two products: FCR-N and FCR-D, corresponding to normal operation (small frequency deviations) and disturbance conditions (larger deviations), respectively. For FCR-N, the response is currently symmetric. However, Energinet and Svenska kraftnät are assessing a potential split into separate upward and downward products \cite{Energinet_survey_nodate}. Similar to FCR cooperation, there is a relatively small domestic FCR-N capacity defined by Denmark’s generation share in the Nordic system. The response speed requirements are less stringent than those in FCR Cooperation, and shorter provision periods are available in the daily auction. In contrast to FCR Cooperation, both availability and energy delivery are remunerated, with energy settled based on the hourly average mFRR price on a dual price basis. On the other hand, FCR-D, is asymmetric and associated with larger capacity requirements, typically determined by the outage of major generation units or lines. Its response speed requirements are significantly more stringent than those of FCR Cooperation and FCR-N, with the majority of the response required within seven seconds. Similar to the FCR Cooperation in DK1, the delivered energy through FCR-D is settled as imbalances.

\section{Discussion on Gaps and Challenges} \label{sec:Gaps}

While certain technical gaps and challenges have been briefly introduced in the literature reviewed in Section I from the perspectives of academia and TSOs, this section elaborates on those particularly relevant to HVDC-OWPP developers, as summarized in Table.~\ref{tab:gaps_challenges_summary}.

\newcommand{\ccell}[2]{\adjustbox{valign=c}{\parbox{#1}{\raggedright#2}}}

\begin{table*}[h!]
\centering
\caption{Gaps and challenges in fast frequency service provision from HVDC-OWPPs}
\label{tab:gaps_challenges_summary}
\footnotesize
\setlength{\tabcolsep}{3pt}
\renewcommand{\arraystretch}{1.2}
\begin{tabular}{@{}p{0.11\textwidth}p{0.12\textwidth}p{0.35\textwidth}p{0.35\textwidth}@{}}
\toprule
\textbf{Services} & & \textbf{Gaps} & \textbf{Challenges} \\

\midrule

\multirow{3}{0.11\textwidth}{Inertia}
& \ccell{0.12\textwidth}{Grid-Code Obligations}
& \ccell{0.35\textwidth}{\begin{itemize}[nosep,leftmargin=*]
    \item Benchmark specification of $T_{\text{M,PPM}}$
    \item Remuneration framework
\end{itemize}}
& \ccell{0.35\textwidth}{\begin{itemize}[nosep,leftmargin=*]
    \item Extra technical complexity and cost due to adjunctive functions
\end{itemize}} \\[8pt]

& \ccell{0.12\textwidth}{Commercial Products} 
& \ccell{0.35\textwidth}{\begin{itemize}[nosep,leftmargin=*]
    \item Stacking and coordination policy with other (fast frequency) services
\end{itemize}}
& \ccell{0.35\textwidth}{\begin{itemize}[nosep,leftmargin=*]
    \item Long-term commitments \& high availability
    \item Remuneration remains unattractive
\end{itemize}}\\[8pt]

& \ccell{0.12\textwidth}{In Common} 
& \ccell{0.35\textwidth}{\begin{itemize}[nosep,leftmargin=*]
    \item Benchmark acceptance criteria of dynamic performance
    \item Long-term impact on components' lifetime
\end{itemize}} 
& \ccell{0.35\textwidth}{\begin{itemize}[nosep,leftmargin=*]
    \item Asymmetric provision
    \item Optimization on the operational cost
\end{itemize}}\\

\midrule

FFR & \ccell{0.12\textwidth}{Commercial Products} 
& \ccell{0.35\textwidth}{\begin{itemize}[nosep,leftmargin=*]
    \item Downward product
\end{itemize}}
& \ccell{0.35\textwidth}{\begin{itemize}[nosep,leftmargin=*]
    \item Long forecast horizon and settlement period
    \item The demand and price in the market remain limited
\end{itemize}}\\

\midrule

\multirow{2}{0.11\textwidth}{FCR}
& \ccell{0.12\textwidth}{Grid-Code Obligations} 
& \ccell{0.35\textwidth}{\begin{itemize}[nosep,leftmargin=*]
    \item Long-term Impact on components' lifetime under higher response speed requirement and tighter combined effect of insensitivity and deadband
\end{itemize}}
& \ccell{0.35\textwidth}{\begin{itemize}[nosep,leftmargin=*]
    \item Higher response speed requirements
    \item Tighter constraints on the combined effect of insensitivity and deadband for FSM
\end{itemize}}\\[32pt]

& \ccell{0.12\textwidth}{Commercial Products} 
& \ccell{0.35\textwidth}{\begin{itemize}[nosep,leftmargin=*]
    \item Long-term Impact on components' lifetime under higher response speed requirement and diverse frequency response ranges
\end{itemize}}
& \ccell{0.35\textwidth}{\begin{itemize}[nosep,leftmargin=*]
    \item Emerging services with diverse frequency ranges
    \item Emerging services with higher response speed requirement
\end{itemize}}\\[8pt]

\midrule
\ccell{0.12\textwidth}{In General} & 
& \ccell{0.35\textwidth}{\begin{itemize}[nosep,leftmargin=*]
    \item Overall requirements on the whole HVDC-OWPP system
\end{itemize}}
& \ccell{0.35\textwidth}{\begin{itemize}[nosep,leftmargin=*]
    \item Coordination of the whole HVDC-OWPP system
    \item Limited Capacity for Fast Response
    \item Stacking and coordination among services
\end{itemize}}\\
\bottomrule
\end{tabular}
\end{table*}

\subsection{Inertia}

\subsubsection{Grid-code obligations for Inertia}

As part of the broader GFM capability requirements, inertia represents only one of several required functionalities. Additional capabilities, including phase jump response, damping support, islanded operation, and other dynamic stability functions, must also be implemented to ensure reliable system operation. The integration of these advanced functionalities significantly increases the technical complexity of converter control systems and their coordination with wind turbine generators. Consequently, meeting these expanded requirements is expected to increase both investment and operational costs for project developers.

While the specific $T_{\text{M,PPM}}$ must be negotiated between the TSO and HVDC-OWPP developer within the inherent energy storage capability, a benchmark of $T_{\text{M,PPM}}$ is not defined yet. The corresponding uncertainty may introduce additional costs and potentially delay the development process. At present, most commercial WTG products are based on GFL control, and retrofitting these turbines with GFM capability entails additional costs. Furthermore, supplementary energy storage or other GFM-supporting equipment may need to be installed to deliver the level of inertia specified by the TSO. Such requirements may necessitate turbine redesign, re-certification, and additional compliance verification. Consequently, turbine procurement, compliance testing, and commissioning processes may experience further delays.

Moreover, existing technical specifications remain insufficiently detailed for early-stage assessment of grid code compliance and market participation, posing practical challenges for developers. In Great Britain, the acceptance criteria are not publicly available, complicating early-stage feasibility analyses. In Germany, the reference behavior and the derived acceptance envelope curves used for compliance verification can only be generated and issued by the relevant TSO, which limits the ability of developers to perform independent pre-assessments. Although a reference implementation of GFM control has been provided in the recent ENTSO-E guidelines \cite[Appendix B]{ENTSOE2025GridFormingPPM} and offers some guidance, it does not fully substitute for the official verification criteria. Therefore, more comprehensive and publicly accessible specifications and acceptance criteria are expected to facilitate early-stage evaluation and project development.

During operation, several gaps remain regarding the practical implementation of inertia provision. First, the remuneration framework for mandatory inertia provision remains uncertain, particularly in the German case, regarding potential imbalance costs associated with rotor speed recovery following inertia response. In addition, the interaction between inertia provision and the delivery of other ancillary services is not yet clearly defined, which may complicate operational strategies for HVDC-OWPPs. Furthermore, the long-term impact of frequent inertia provision on component lifetime, especially for power electronic converters and turbine drivetrain components, has not yet been fully evaluated. These uncertainties highlight the need for further assessment of both the economic and technical implications of inertia provision during normal operation.

\subsubsection{Commercial Products for Inertia}
The current market arrangements and product designs remain relatively inflexible, which makes HVDC-OWPP developers cautious about participating. In both Great Britain and Germany, the emerging inertia markets involve long-term commitments, while settlement intervals are typically 15 to 30 min.This creates a mismatch, as wind power availability cannot be predicted at such temporal resolution over long horizons. In Great Britain, inertia provision is required to be bi-directional but there is no mandatory contribution. Hence, inertia within inherent capability (i.e., rotor kinetic energy) could be used for relatively low-risk participation. Nevertheless, developers must consider the opportunity cost of potential curtailment needed to reserve power for inertia provision, as well as possible imbalance costs associated with rotor speed recovery, which complicates the economic assessment. In Germany, more flexible products allowing provision in either direction are available. However, since the inherent capability will be taken to meet the mandatory requirements, it can be challenging to achieve the high availability requirement of commercial products beyond inherent capability. These limitations indicate a need for more flexible market designs and products, such as shorter-term procurement mechanisms, such as the day-ahead (D-1) market currently under development in Great Britain.

Since the technical specifications and compliance requirements for commercial inertia services largely mirror those in mandatory requirements, many of the associated gaps and challenges are similar to those discussed for mandatory requirements, spanning stages from initial assessment and compliance verification to operational implementation. In particular, the non-public compliance requirements in the British market appear to be sufficiently stringent that, to date, no non-synchronous GFM unit has successfully passed the technical assessment. In addition, the assessment of service availability can be particularly challenging. For instance, in Germany, availability is evaluated based on minute-resolution measurements. As a result, HVDC-OWPP operators may need to maintain additional operational margins to mitigate the impact of wind power fluctuations and the effects of concurrent inertia or other ancillary service provisions, ensuring that compliance with the availability criteria can be reliably demonstrated.

Finally, as inertia is an emerging ancillary service, the current remuneration mechanisms in both Great Britain and Germany are designed to provide a certain degree of revenue certainty. In Great Britain, the pay-as-bid mechanism, together with the indicative Payment Calculator, allows developers to estimate potential revenues from participation. In contrast, the German instantaneous reserve market adopts a fixed-price availability payment, offering even greater revenue certainty. However, while revenues in the British market depend on competitive bidding, the fixed-price structure in Germany may not be sufficiently attractive to generate adequate profits after accounting for various costs, including the direct costs of implementing grid-forming (GFM) functionality, opportunity costs associated with reserving power capacity for inertia provision, potential imbalance costs during power recovery, and depreciation of wind turbine generator hardware. Consequently, the current remuneration structures may not yet provide sufficient economic incentives for large-scale participation.

\subsubsection{Asymmetric Inertia Provision}
\begin{figure}[htbp]
    \centering
    \includegraphics[width=0.6\textwidth]{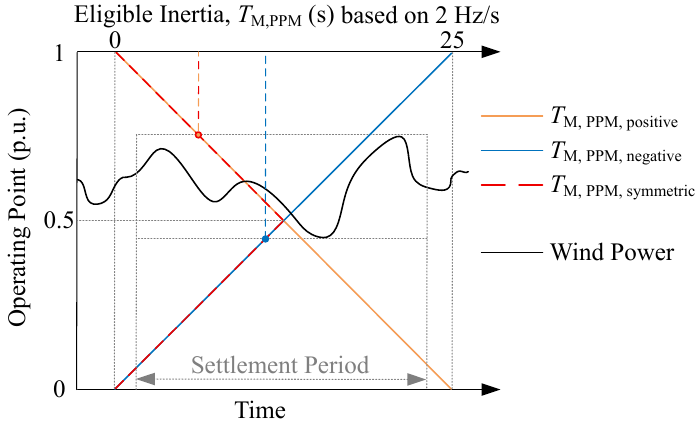} 
    \caption{Eligible inertia under variable wind power in the German context.}
    \label{fig:AsymInertia}
\end{figure}

Asymmetric inertia provision is required in NC 2.0 for Type-B generation unit (only negative inertia to over-frequency event) and is also defined for German instantaneous reserve market. In particular, the availability of instantaneous reserve is evaluated based on the available converter headroom for positive inertia and footroom for negative inertia. Consequently, the eligible positive, negative, and symmetric inertia may differ substantially, depending on the maximum and minimum active power output during the settlement period, as illustrated in Figure~\ref{fig:AsymInertia}.

At present, mainstream GFM control schemes provide only symmetric inertia \cite{10981620}. Such schemes may therefore be unable to fully exploit the available inertia capability under variable wind production, as indicated Figure~\ref{fig:AsymInertia}. Only few studies, such as \cite{Heid2022Asymmetric,11540169,Rehman2026Asymmetric}, have investigated asymmetric inertia provision. However, control approaches capable of fully uni-directional inertia provision remain insufficiently established.

\subsubsection{Curtailment or Imbalance for Inertia}
\begin{figure}[t!]
    \centering
    \includegraphics[width=\textwidth]{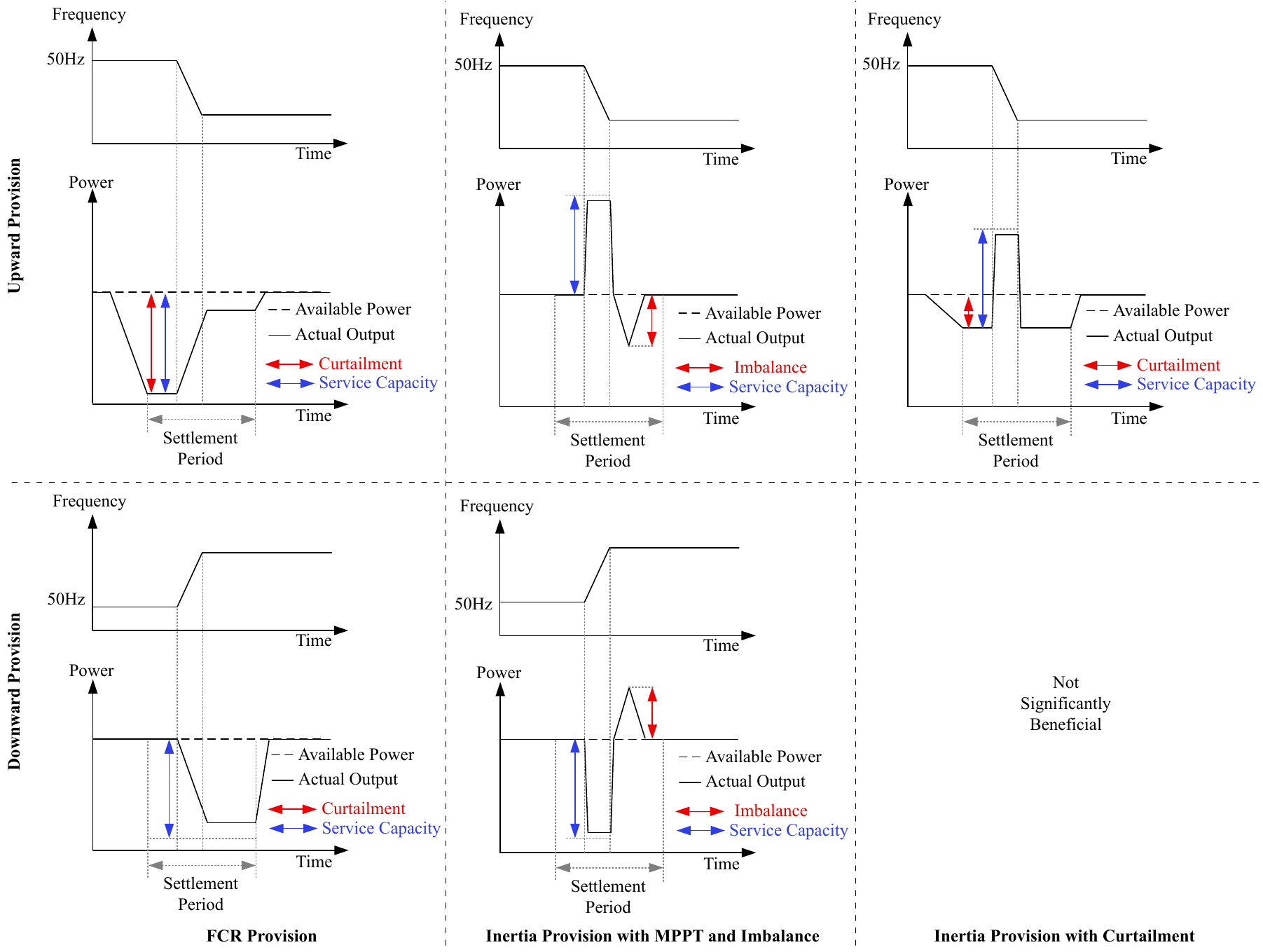} 
    \caption{Provision of FCR and inertia with curtailment or imbalance.}
    \label{fig:CurtialorImbalance}
\end{figure}

Regarding upward provision, sustainable frequency reserves such as FCR require pre-curtailment at least equal to the service capacity prior to delivery. In contrast, inertia provision can be achieved based on either pre-curtailment or post-imbalance (post-recovery), and in both cases the required power offset can be less than the service capacity, as shown in Figure~\ref{fig:CurtialorImbalance}. Therefore, the leverage between the pre-curtailment or post-imbalance and the service capacity confers an operational advantage to inertia provision. However, the minimum pre-curtailment or post-imbalance required for inertia provision under GFM control remains to be evaluated.

Regarding downward provision, sustainable frequency reserves such as FCR no longer require pre-curtailment. However, inertia provision still necessitates a post-imbalance margin, owing to the mismatch between the timescales of inertia delivery and rotor speed regulation by pitch control, as shown in Figure~\ref{fig:CurtialorImbalance}. Alternatively, pre-curtailment in this case is of limited benefit. In this context, the post-imbalance requirement constitutes an operational disadvantage. Nevertheless, the significance of this disadvantage may be substantially reduced through optimization, and its characterization likewise remains a subject for future investigation.

\subsection{Fast Frequency Reserve}
The day-ahead FFR market in DK2 features a shorter forecast horizon than the long-term commercial inertia markets currently in place, which improves its operational alignment with variable wind power generation. Nevertheless, the horizon may still be relatively long for HVDC-OWPPs. In addition, the market’s 1-h trading interval remains comparatively coarse, while the current level of wind power forecasting accuracy cannot reliably support participation at this temporal resolution. 

On the positive side, the technical specifications and compliance requirements for FFR provision are clearly and comprehensively defined, allowing potential participants to conduct a thorough pre-assessment of their capability to deliver the service. However, several design aspects may limit the economic attractiveness for HVDC-OWPPs. In particular, the service is restricted to upward reserve provision, which reduces operational flexibility. To maintain upward reserve capacity, HVDC-OWPPs typically need to operate in a curtailed state to create headroom or perform rotor speed recovery following the reserve activation. Curtailment leads to opportunity costs associated with lost energy production, while rotor speed recovery may introduce imbalance costs if the resulting generation deviates from the scheduled output. When combined with equipment depreciation and other operational costs, these factors may reduce the overall economic viability of participation if the remuneration does not adequately compensate the associated costs. 

Although the pay-as-clear pricing mechanism applied in the FFR market is generally considered more attractive than the pay-as-bid or fixed-price remuneration schemes used in the inertia markets, historical data suggest that the average FFR demand and price have typically been less competitive than those observed in other ancillary service markets, such as mFRR \cite{EnerginetPricesProcurementProjections}.

\subsection{Frequency Containment Reserve}

\subsubsection{Grid-code Obligations for Frequency Containment Reserve}

The forthcoming tightening of response speed requirements may pose a challenge. Under NC 1.0, baseline response speed requirements are defined, while the IGD recommends more stringent criteria for power park modules compared to synchronous generating units. In current national implementations (e.g., in DE, GB, DK that this paper concerns), these IGD specifications are often relaxed to some extent. In turn, NC 2.0 is expected to adapt and formalize these more stringent IGD criteria, leading to stricter compliance requirements.

In addition, NC 2.0 introduces tighter constraints on the combined effect of insensitivity and deadband for FSM, which may present further challenges. Increased sensitivity to frequency deviations is likely to result in more frequent activation and operational duty cycles, potentially leading to increased mechanical wear and fatigue, as well as increased energy delivery in respective directions. Moreover, control systems must support high-resolution frequency measurement and fast small-signal response. Depending on the specific implementation of FSM requirements, a comprehensive assessment will be needed on the associated costs, including increased activation frequency, equipment stress, and control complexity.

\subsubsection{Commercial Products for Frequency Containment Reserve}
 Emerging FCR markets are characterized by new response requirements in both the frequency and time domains, thereby introducing opportunities as well as challenges. 
 
 In the frequency domain, conventional FCR products, such as FCR Cooperation, typically require a proportional response to frequency deviations within the steady-state range (i.e., $\pm 200 \text{mHz}$) with minor deadband. Emerging FCR products retain this proportional behavior but may target different frequency ranges (e.g., FCR-D operating within deviations of $\pm [100, 500] \text{mHz}$) or adopt piecewise response characteristics (e.g., DC and DM in Great Britain). These modified frequency characteristics may lead to different energy-utilization patterns and, consequently, affect the operational costs of reserve provision. 
 
 In the time domain, emerging FCR products generally impose substantially faster response requirements, typically within 1–10 s, reflecting the needs of converter-dominated power systems and the characteristics of renewable generation. Faster active-power responses may increase mechanical loading, potentially leading to higher maintenance costs and reduced component lifetime. Moreover, the available capacity for rapid response may become more constrained as shorter delivery times are required. Nevertheless, given the stringent speed requirements and the relatively recent introduction of these markets, early certification and participation can provide a competitive advantage.

\subsection{General Gaps and Challenges}
In this subsection, the gaps and challenges apply to all the fast frequency services of inertia, FFR, and FCR.

\subsubsection{Overall Requirements and Coordination of the whole HVDC-OWPP System}

At present, although grid codes typically specify requirements for transferring onshore frequency-variation signals to the offshore side, frequency-response requirements are generally defined separately for the HVDC system and the OWPP, with each assessed primarily at its own point of connection. However, studies in \cite{saborio2020communication,jiang2023novel,
Xu2026HolisticGFM} have demonstrated that coordinated control between the HVDC system and the OWPP can improve the performance and effectiveness of frequency response at both the onshore and offshore connection points. Owing to the dynamic interactions between the HVDC system and the OWPP, the frequency-response behavior observed in standalone OWPP compliance tests may differ from that achieved during actual service provision through an integrated HVDC-OWPP system. Therefore, frequency response requirements should be specified at the level of the overall HVDC-OWPP system, rather than solely at the individual subsystem level. Such a system-level specification would explicitly account for HVDC–OWPP interactions, promote coherent response performance, and facilitate the development of coordinated control strategies to enhance the technical competitiveness of HVDC-OWPPs in frequency response service provision.

\subsubsection{Limited Capacity for Fast Response}

\begin{figure}[htbp]
    \centering
    \includegraphics[width=0.8\textwidth]{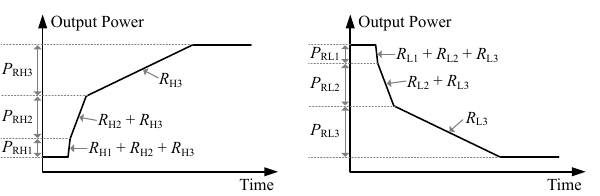} 
    \caption{An example of ramp rate limits of a (GFM) WTG.}
    \label{fig:RampRateLimit}
\end{figure}

Although the power electronic converters of WTGs are capable of delivering the rapid response required for inertia and FCR provision, such fast response can impose significant mechanical stress on WTG components (e.g., rotor, drive train, etc.), thereby reducing their operational lifetime \cite{bjorn2022power,heidenreich2015wind}. To mitigate this issue, a practical approach is to impose piecewise ramp rate limits on the output power. An example is illustrated in Figure~\ref{fig:RampRateLimit} and formalized in \eqref{RampRateLimit}, where $P$ (p.u.) and $P_{\text{lim}}$ (p.u.) denote the output power before and after ramp rate limits, respectively; $i$ is the control cycle index; $t$ (s) is the time instant; $\Delta t$ (s) is the time step of a control cycle; $r_{\text{H}}$ (p.u./s) and $r_{\text{L}}$ (p.u./s) are the high and low ramp rate limits, respectively; $j$ is the index of ramp rate limit sections; $n$ is the total number of ramp rate sections; $R_{\text{H}}$ (p.u./s, positive) and $R_{\text{L}}$ (p.u./s, negative) are the limit values of the high and low ramp rates; $P_{\text{RH}}$ (p.u., positive) and $P_{\text{RL}}$ (p.u., negative) are the capacity ranges of the high and low ramp rate limits; $k$ is the index of the cumulative ramp rate sections. Typical parameter settings are summarized in Table~\ref{tab:RampRateLimit}. Consequently, the available capacity for fast response is generally very limited, constraining the provision of fast frequency services from HVDC-OWPPs.

\begin{subequations}
\label{RampRateLimit}
\begin{align}
        &P_{\text{lim}}(i) = 
    \left\{\begin{array}{ll}
        \Delta t \cdot \sum_{j = 1}^{n} r_{\text{H}j}(i) + P_{\text{lim}}(i-1) 
        & \text{, if } \frac{P(i)-P_{\text{lim}}(i-1)}{t(i)-t(i-1)} \geq \sum_{j = 1}^{n} r_{\text{H}j}(i)\\ 
        P(i) 
        & \text{, if } \sum_{j = 1}^{n} r_{\text{L}j}(i) < \frac{P(i)-P_{\text{lim}}(i-1)}{t(i)-t(i-1)} < \sum_{j = 1}^{n} r_{\text{H}j}(i) \\ 
        \Delta t \cdot \sum_{j = 1}^{n} r_{\text{L}j}(i) + P_{\text{lim}}(i-1) 
        & \text{, if } \frac{P(i)-P_{\text{lim}}(i-1)}{t(i)-t(i-1)} \leq \sum_{j = 1}^{n} r_{\text{L}j}(i)
    \end{array}\right.
    \\
    &r_{\text{H}j}(i) = 
        \left\{\begin{array}{ll}
        R_{\text{H}j}
        & \text{, if} \, P_{\text{lim}}(i-1) - P_{\text{lim}}(i-1- \lfloor \frac{1}{\Delta t}\sum_{k = 1}^{j} \frac{\Delta P_{\text{RH}k}}{R_{\text{H}k}} \rceil) < \Delta P_{\text{RH}j}
        \\
        0
        & \text{, if} \, P_{\text{lim}}(i-1) - P_{\text{lim}}(i-1- \lfloor \frac{1}{\Delta t}\sum_{k = 1}^{j} \frac{\Delta P_{\text{RH}k}}{R_{\text{H}k}} \rceil) \geq \Delta P_{\text{RH}j}
            \end{array}\right.
            \\
     &r_{\text{L}j}(i) = 
        \left\{\begin{array}{ll}
        R_{\text{L}j}
        & \text{, if} \, P_{\text{lim}}(i-1) - P_{\text{lim}}(i-1- \lfloor \frac{1}{\Delta t}\sum_{k = 1}^{j} \frac{\Delta P_{\text{RL}k}}{R_{\text{L}k}} \rceil) > \Delta P_{\text{RL}j}
        \\
        0
        & \text{, if} \, P_{\text{lim}}(i-1) - P_{\text{lim}}(i-1- \lfloor \frac{1}{\Delta t}\sum_{k = 1}^{j} \frac{\Delta P_{\text{RL}k}}{R_{\text{L}k}} \rceil) \leq \Delta P_{\text{RL}j}
            \end{array}\right.
\end{align}
\end{subequations}

\begin{table}[!t]
\caption{Example settings of ramp rate limits.\label{tab:RampRateLimit}}
\centering
\begin{tabular}{ c c c c }
\hline
\multicolumn{2}{c}{\textbf{Ramp Rate Limit (p.u./s)}} & \multicolumn{2}{c}{\textbf{Corresponding Capacity (p.u.)}}\\
\hline
$R_{\text{H}1}$ & 0.2 & $P_{\text{RH}1}$ & 0.1 \\

$R_{\text{H}2}$ & 0.07 & $P_{\text{RH}2}$ & 0.33 \\

$R_{\text{H}3}$ & 0.03 & $P_{\text{RH}3}$ & 0.57 \\

$R_{\text{L}1}$ & -0.2 & $P_{\text{RL}1}$ & -0.1 \\

$R_{\text{L}2}$ & -0.07 & $P_{\text{RL}2}$ & -0.33 \\

$R_{\text{L}3}$ & -0.03 & $P_{\text{RL}3}$ & -0.57 \\
\hline
\end{tabular}
\end{table}

\subsubsection{Stacking and Coordination among Services}

\begin{figure*}[h!]
\centering
\subfloat[]
{\includegraphics[width=0.32\textwidth]{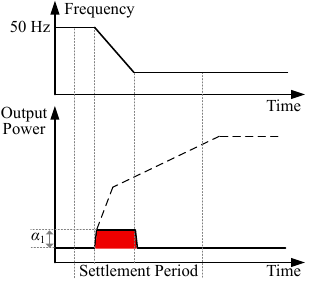}%
\label{fig:CoordinationSub1}}
\hfil
\subfloat[]
{\includegraphics[width=0.32\textwidth]{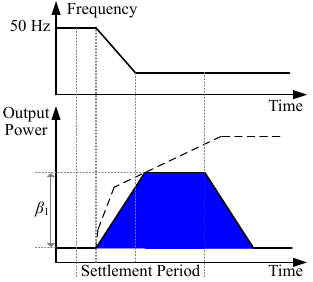}%
\label{fig:CoordinationSub2}}
\hfil
\subfloat[]
{\includegraphics[width=0.32\textwidth]{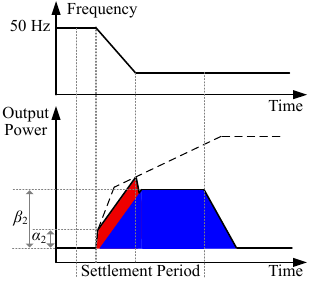}%
\label{fig:CoordinationSub3}}
\caption{Coordination among services. (a) Maximum inertia provision; (b) Maximum FCR provision; (c) Stacking inertia and FCR.}
\label{fig_Coordination&Stacking}
\end{figure*}

As a direct consequence of the limited fast-response capacity, faster frequency services are subject to increasingly constrained provision capacity. As shown in Figure~\ref{fig:CoordinationSub1} and \ref{fig:CoordinationSub2}, the maximum available FCR provision substantially exceeds the maximum inertia provision, since only the initial, fastest ramp can satisfy the speed requirements of inertia response, whereas FCR provision is compatible with slower ramp rates. As a result, faster frequency services utilize a smaller fraction of the HVDC-OWPP's ramping capability. By stacking inertia and FCR services together, the ramping capability of the HVDC-OWPP can be fully exploited, as depicted in Figure~\ref{fig:CoordinationSub3}. However, the combined service capacity ($\alpha_{2} + \beta_{2}$) may be greater or less than the maximum provision capacity of the slowest individual service ($\beta_{1}$), depending on the permissible time delay, the acceptable response deviation, and the specific speed requirements of each service. Moreover, service pricing significantly influences the stacking strategy, making the optimal coordination highly market- and country-specific. Nevertheless, a key implication is that accurate modeling of fast transient dynamics is essential, substantially increasing the optimization complexity.

\section{Conclusion}
\label{sec:Conc}
The evolving regulatory and market frameworks for inertia, FFR, and FCR provision present both significant opportunities and substantial challenges for HVDC-OWPPs. 

Across Europe, the transition from optional to future mandatory inertia capabilities reflects the increasing need for converter-dominated power systems to support system stability. However, these developments also introduce considerable technical complexity, higher investment and operational costs, and increased uncertainty related to mandatory baseline, compliance verification, operational impacts, and remuneration mechanisms. In particular, the lack of transparent acceptance criteria and the uncertain interaction between ancillary services complicate both project development and operational planning.

Commercial inertia and FFR services remain constrained by product designs that are not yet fully compatible with the variable nature of offshore wind generation. The long procurement horizons, coarse temporal granularity, and strict availability requirements of inertia services, as well as the necessary curtailment for FFR, reduce the economic attractiveness. Although remuneration schemes provide varying degrees of revenue certainty, they may still be insufficient to offset implementation costs, opportunity costs, imbalance exposure, and equipment degradation.

On the other hand, the development of emerging FCR products appears increasingly favorable for OWPP participation. Faster response capabilities of converter-based resources align well with evolving system needs, while trends toward asymmetric provision, shorter procurement horizons, reduced delivery granularity, and near real-time markets improve compatibility with wind power variability and forecasting limitations. Nevertheless, tightening response requirements and increased activation sensitivity may introduce additional technical and operational burdens. 

HVDC-OWPPs face three key technical gaps/challenges in fast frequency service provision. Mechanical stress constraints on WTG components necessitate ramp rate limits, restricting available fast-response capacity. Stacking multiple fast frequency services such as inertia and FCR can better utilize ramping capability, but optimal coordination is market-specific and requires accurate fast-dynamic modeling. Additionally, the energy offset requirements differ asymmetrically between inertia and FCR for upward and downward reserves, and their optimal characterization under GFM control remains an open research question.

Overall, future large-scale participation of HVDC-OWPPs in fast frequency service markets will depend on the continued evolution of market designs, transparent compliance frameworks, remuneration mechanisms, and operation optimization that adequately reflect the technical capabilities, operational constraints, and economic risks of offshore wind power.

\section{Acknowledgment}
This work was supported by the ADOreD project of the European Union’s Horizon Europe Research and Innovation Program under the Marie Skłodowska-Curie Grant 101073554.

\bibliographystyle{model3-num-names}
\bibliography{refs.bib} 

@legislation{noauthor_directive_2023,
	title = {Directive ({EU}) 2023/2413 of the European Parliament and of the Council of 18 October 2023 amending Directive ({EU}) 2018/2001, Regulation ({EU}) 2018/1999 and Directive 98/70/{EC} as regards the promotion of energy from renewable sources, and repealing Council Directive ({EU}) 2015/652},
	url = {http://data.europa.eu/eli/dir/2023/2413/oj/eng},
	urldate = {2025-02-27},
	date = {2023-10-18},
	langid = {english},
    year = {n.d.},
    note = {[accessed 8 July 2026]},
}

@misc{noauthor_renewable_nodate,
	title = {Renewable energy targets - European Commission},
	url = {https://energy.ec.europa.eu/topics/renewable-energy/renewable-energy-directive-targets-and-rules/renewable-energy-targets_en},
	urldate = {2025-02-27},
	langid = {english},
    year         = {n.d.},
    note = {[accessed 8 July 2026]},
}

@misc{noauthor_offshore_nodate,
	title = {Offshore renewable energy},
	url = {https://energy.ec.europa.eu/topics/renewable-energy/offshore-renewable-energy_en},
	urldate = {2025-02-28},
	langid = {english},
    year = {n.d.},
    note = {[accessed 8 July 2026]},
}

@legislation{noauthor_communication_2023,
	title = {{COMMUNICATION} {FROM} {THE} {COMMISSION} {TO} {THE} {EUROPEAN} {PARLIAMENT}, {THE} {COUNCIL}, {THE} {EUROPEAN} {ECONOMIC} {AND} {SOCIAL} {COMMITTEE} {AND} {THE} {COMMITTEE} {OF} {THE} {REGIONS} European Wind Power Action Plan},
	url = {https://eur-lex.europa.eu/legal-content/EN/TXT/?uri=CELEX%3A52023DC0669&qid=1702455143415},
	urldate = {2025-02-28},
	date = {2023},
	langid = {english},
    year = {n.d.},
note = {[accessed 8 July 2026]},
}

@misc{noauthor_north_nodate,
	title = {North Sea Energy Island},
	url = {https://northseaenergyisland.dk/en},
	urldate = {2025-02-28},
	langid = {english},
    year = {n.d.},
note = {[accessed 8 July 2026]},
}

@misc{noauthor_energy_nodate,
	title = {Energy Island Bornholm},
	url = {https://www.energiobornholm.dk/en},
	urldate = {2025-02-28},
	langid = {english},
    year = {n.d.},
note = {[accessed 8 July 2026]},
}

@article{yang_critical_2022,
  author  = {Yang, Bo and Liu, Bingqiang and Zhou, Hongyu and Wang, Jingbo and Yao, Wei and Wu, Shaocong and Shu, Hongchun and Ren, Yaxing},
  title   = {A critical survey of technologies of large offshore wind farm integration: summary, advances, and perspectives},
  journal = {Prot Control Mod Power Syst},
  year    = {2022},
  volume  = {7},
  pages   = {17},
  doi     = {10.1186/s41601-022-00239-w},
  url     = {https://doi.org/10.1186/s41601-022-00239-w}
}

@article{rancilio_ancillary_2022,
author  = {Rancilio, G. and Rossi, A. and Falabretti, D. and Galliani, A. and Merlo, M.},
  title   = {Ancillary services markets in {Europe}: evolution and regulatory trade-offs},
  journal = {Renew Sustain Energy Rev},
  year    = {2022},
  volume  = {154},
  pages   = {111850},
  doi     = {10.1016/j.rser.2021.111850},
  url     = {https://www.sciencedirect.com/science/article/pii/S1364032121011175}
}

@report{erraia_offshore_2023,
	location = {Oslo},
	title = {{OFFSHORE} {WIND} {SUBSIDIES} {IN} {THE} {EU}, {NORWAY}, {AND} {THE} {US}},
	url = {https://menon.no/en/projects/offshore-wind-subsidies-in-the-eu-norway-and-the-us},
	number = {{MENON} {PUBLICATION} {NO}. 51/2023},
	institution = {Menon Economics},
	author = {Erraia, Jonas and Foseid, Henrik and Śpiewanowski, Piotr and Winje, Even and Wahl, Einar},
	year = {2023},
}

@legislation{noauthor_commission_2020,
	title = {{COMMISSION} {STAFF} {WORKING} {DOCUMENT} Accompanying the document {COMMUNICATION} {FROM} {THE} {COMMISSION} {TO} {THE} {EUROPEAN} {PARLIAMENT}, {THE} {COUNCIL}, {THE} {EUROPEAN} {ECONOMIC} {AND} {SOCIAL} {COMMITTEE} {AND} {THE} {COMMITTEE} {OF} {THE} {REGIONS} An {EU} strategy to harness the potential of offshore renewable energy for a climate neutral future},
	url = {https://eur-lex.europa.eu/legal-content/EN/TXT/?uri=SWD:2020:273:FIN},
	year = {2020-11-19},
	date = {2020},
	langid = {english},
}

@article{jansen_offshore_2020,
	author  = {Jansen, Malte and Staffell, Iain and Kitzing, Lena and Quoilin, Sylvain and Wiggelinkhuizen, Edwin and Bulder, Bernard and Riepin, Iegor and M{\"u}sgens, Felix},
  title   = {Offshore wind competitiveness in mature markets without subsidy},
  journal = {Nat Energy},
  year    = {2020},
  volume  = {5},
  number  = {8},
  pages   = {614--622},
  doi     = {10.1038/s41560-020-0661-2},
  url     = {https://www.nature.com/articles/s41560-020-0661-2}
}

@report{noauthor_new_2018,
    author = {{Aurora Energy Research Ltd.}},
	title = {The new economics of offshore wind},
	url = {https://auroraer.com/wp-content/uploads/2021/04/The-new-economics-of-offshore-wind.-Aurora-Energy-Research-Report.pdf},
	year = {2018},
}

@article{cole_critical_2023,
author  = {Cole, Matthew and Campos-Gaona, David and Stock, Adam and Nedd, Marcel},
  title   = {A critical review of current and future options for wind farm participation in ancillary service provision},
  journal = {Energies},
  year    = {2023},
  volume  = {16},
  number  = {3},
  pages   = {1324},
  doi     = {10.3390/en16031324},
  url     = {https://www.mdpi.com/1996-1073/16/3/1324}
}

@legislation{noauthor_commission_2016_hvdc,
	title = {Commission Regulation ({EU}) 2016/1447 of 26 August 2016 establishing a network code on requirements for grid connection of high voltage direct current systems and direct current-connected power park modules (Text with {EEA} relevance)},
	url = {http://data.europa.eu/eli/reg/2016/1447/oj/eng},
	urldate = {2024-01-18},
	year = {2016},
	langid = {english},
	note = {Legislative Body: {COM}},
}

@article{wu2024grid,
  title={Grid integration of offshore wind power: standards, control, power quality and transmission},
  author={Wu, Dan and Seo, Gab-Su and Xu, Lie and Su, Chi and Kocewiak, {\L}ukasz and Sun, Yin and Qin, Zian},
  journal={IEEE Open Journal of Power Electronics},
  year={2024},
  publisher={IEEE}
}

@article{viola_ancillary_2024,
	author  = {Viola, L. and Mohammadi, S. and Dotta, D. and Hesamzadeh, M. R. and Baldick, R. and Flynn, D.},
  title   = {Ancillary services in power system transition toward a 100\% non-fossil future: market design challenges in the {United States} and {Europe}},
  journal = {Electr Power Syst Res},
  year    = {2024},
  volume  = {236},
  pages   = {110885},
  doi     = {10.1016/j.epsr.2024.110885}
}

@article{li_review_2023,
	author  = {Li, Le and Zhu, Donghai and Zou, Xudong and Hu, Jiabing and Kang, Yong and Guerrero, Josep M.},
  title   = {Review of frequency regulation requirements for wind power plants in international grid codes},
  journal = {Renew Sustain Energy Rev},
  year    = {2023},
  volume  = {187},
  pages   = {113731},
  doi     = {10.1016/j.rser.2023.113731},
  url     = {https://www.sciencedirect.com/science/article/pii/S1364032123005889}
}

@legislation{noauthor_commission_2016_rfg,
	title = {Commission Regulation ({EU}) 2016/631 of 14 April 2016 establishing a network code on requirements for grid connection of generators (Text with {EEA} relevance)},
	url = {http://data.europa.eu/eli/reg/2016/631/oj/eng},
	urldate = {2024-01-18},
	year = {2016},
	langid = {english},
}

@techreport{vde4131,
    author = {VDE FNN},
    title = {VDE-AR-N 4131 Technical requirements for grid connection of high voltage direct current systems and direct current-connected power park modules (TCR HVDC)},
    institution = {VDE FNN},
    year = {2019}
}

@techreport{NAR,
    author      = {{TenneT TSO GmbH}},
  title       = {{Grid Connection Requirements -- High and Extra-High Voltage}},
  institution = {{TenneT TSO GmbH}},
  address     = {Bayreuth, Germany},
  year        = {2026},
  note        = {Version as of 31 March 2026}
}

@misc{noauthor_ancillary_2025_energinet,
	title = {Ancillary Services To Be Delivered In Denmark - Tender Conditions},
	url = {https://en.energinet.dk/electricity/ancillary-services/tender-conditions-for-ancillary-services/},
	publisher = {Energinet},
	year = {2025},
}

@misc{ESO_2025_CUSC,
author       = {{National Energy System Operator}},
  title        = {{Connection and Use of System Code (CUSC)}},
  year         = {2026},
  month        = may,
  note         = {Complete CUSC, 29 May 2026},
  url          = {https://www.neso.energy/industry-information/codes/connection-and-use-system-code-cusc/cusc-code-documents},
  organization = {{National Energy System Operator}}
}

@misc{energinet_Guidelines_TR325_2016,
	author      = {{Energinet.dk}},
  title       = {{Technical Regulation 3.2.5 for Wind Power Plants above 11 kW}},
  institution = {{Energinet.dk}},
  number      = {{TR 3.2.5}},
  year        = {2016},
  month       = jul,
  note        = {Revision 4.1; effective from 22 July 2016; expired},
  url         = {https://en.energinet.dk/media/5pbflmly/technical-regulation-325-wind-power-plants-above-11-kw-rev-4_expired.pdf}
}

@misc{ESO_DCDMDR_2024,
	author      = {{National Energy System Operator}},
  title       = {{Dynamic Response Services -- Provider Guidance}},
  institution = {{National Energy System Operator}},
  year        = {2026},
  month       = jun,
  note        = {Version 14; published 2 June 2026},
  url         = {https://www.neso.energy/document/276606/download}
}

@misc{ESO_FCR_procurement_2024,
	author      = {{National Energy System Operator}},
  title       = {{Response Services -- Procurement Rules}},
  institution = {{National Energy System Operator}},
  year        = {2026},
  month       = oct,
  note        = {Version 5; effective 31st March 2026},
  url         = {https://www.neso.energy/document/378246/download}
}

@misc{ESO_FCR_service_term_2024,
	author      = {{National Energy System Operator}},
  title       = {{Response Services -- Service Terms}},
  institution = {{National Energy System Operator}},
  year        = {2025},
  month       = aug,
  note        = {Version 5; effective from 23:00 on 2 September 2025; published 28 August 2025},
  url         = {https://www.neso.energy/document/367526/download}
}

@misc{ESO_SFFR_ServiceTerm_2023,
	 author      = {{National Grid Electricity System Operator Limited}},
  title       = {{Static FFR -- Service Terms}},
  institution = {{National Grid Electricity System Operator Limited}},
  year        = {2023},
  note        = {Version 1.0; effective April 01, 2023},
  url         = {https://www.neso.energy/document/278031/download}
}

@misc{ESO_SFFR_procurement_2023,
	author      = {{National Grid Electricity System Operator Limited}},
  title       = {{Static FFR -- Procurement Rules}},
  institution = {{National Grid Electricity System Operator Limited}},
  year        = {2023},
  month       = mar,
  note        = {Version 1.0; effective April 01, 2023},
  url         = {https://www.neso.energy/document/278036/download}
}

@legislation{ENTSOE_TSO2017/1485_2017,
	title = {Commission Regulation ({EU}) 2017/1485 of 2 August 2017 establishing a guideline on electricity transmission system operation (Text with {EEA} relevance. )},
	url = {http://data.europa.eu/eli/reg/2017/1485/oj/eng},
	urldate = {2024-01-22},
	year = {2017},
	langid = {english},
}

@misc{noauthor_wwwregelleistungnet_nodate,
	title = {www.regelleistung.net {\textgreater} European cooperations {\textgreater} {FCR} Cooperation},
	url = {https://www.regelleistung.net/en-us/European-cooperations/FCR-Cooperation},
	year = {n.d.},
    note = {[accessed: 2026-07-08]},
}

@misc{ENTSOE_FCRcooperation,
	title = {Frequency Containment Reserves},
	url = {https://www.entsoe.eu/network_codes/eb/fcr/#contacts},
	urldate = {2025-04-16},
	langid = {english},
    year = {n.d.},
    note = {[accessed: 2026-07-08]},
}

@techreport{ACER2024HVDCAmendment,
  author      = {{Agency for the Cooperation of Energy Regulators (ACER)}},
  title       = {ACER Recommendation 01--2024 on reasoned proposals for amendments to the network code on requirements for grid connection of high voltage direct current systems and direct current-connected power park modules -- Annex 1: Proposed amended HVDC Regulation},
  year        = {2024},
  institution = {Agency for the Cooperation of Energy Regulators (ACER)},
  number      = {01--2024},
  date        = {2024-12-20}
}

@techreport{ACER2023RfGAmendment,
  author      = {{Agency for the Cooperation of Energy Regulators (ACER)}},
  title       = {ACER Recommendation 03-2023 on reasoned proposals for amendments to the network codes on requirements for grid connection of generators and on demand connection -- Annex 1 – Amended RfG Regulation},
  year        = {2023},
  institution = {Agency for the Cooperation of Energy Regulators (ACER)},
  number      = {03--2023},
  date        = {2023-12-19}
}

@techreport{ENTSOE2025GridFormingPPM,
  author      = {ENTSO-E},
  title       = {Grid forming capability of power park modules: Report on technical requirements},
  year        = {2025},
  institution = {ENTSO-E},
  date        = {2025-10-03}
}

@techreport{VDEFNN2020GFM,
  author       = {{VDE FNN Netztechnik/Netzbetrieb}},
  title        = {{Grid-Forming Behaviour of HVDC Systems and DC-Connected Power Park Modules}},
  institution  = {{VDE Forum Netztechnik/Netzbetrieb (FNN)}},
  type         = {VDE FNN Hinweis},
  number       = {},
  year         = {2020},
  month        = aug,
  note         = {94 pages}
}

@techreport{VDEFNN2025GFM,
  author       = {{VDE FNN Netztechnik/Netzbetrieb}},
  title        = {{Technical Requirements for Grid-Forming Capabilities Including Provision of Inertia: Requirements and Verifications for Grid-Forming Units}},
  institution  = {{VDE Forum Netztechnik/Netzbetrieb (FNN)}},
  type         = {Technical Specification},
  version      = {2.0},
  year         = {2025},
  month        = oct,
  note         = {English version; original German version first published in May 2025}
}

@techreport{NESO_Stability_25_26_ITT_V5,
  author       = {{National Energy System Operator (NESO)}},
  title        = {Mid-Term Stability Market 25/26: Instructions to Tenderers (ITT V5)},
  institution  = {National Energy System Operator},
  year         = {2026},
  month        = jan,
  note         = {Public document}
}

@techreport{NESO_Stability_26_27_ITT_V2,
  author       = {{National Energy System Operator (NESO)}},
  title        = {Mid-Term Stability Market 26/27: Instructions to Tenderers (ITT v2)},
  institution  = {National Energy System Operator},
  year         = {2026},
  note         = {Public document}
}

@techreport{NESO_Stability_27_28_EOI_V1,
  author       = {{National Energy System Operator (NESO)}},
  title        = {Mid-Term (Y-1) Stability Market 27/28: Instructions to Tenderers – Expression of Interest (EOI) and Consultation (v1.0)},
  institution  = {National Energy System Operator},
  year         = {2026},
  month        = jan,
  note         = {Public document}
}

@techreport{NESO_LT2029_ITT_2025,
  title        = {Instructions to Tenderers: Invitation to Tender (ITT) Stage - Long-term 2029: Stability, Voltage and Restoration Services},
  author       = {{National Energy System Operator (NESO)}},
  year         = {2025},
  month        = oct,
  number       = {NESO Tender Reference No: WS2113158887},
  version      = {V4},
  institution  = {National Energy System Operator (NESO)},
  type         = {Instructions to Tenderers},
  url          = {https://www.neso.energy/industry-information/balancing-services/stability-market/long-term-2029-tender},
  note         = {Tender submission deadline extended to 1 May 2026. ITT Pack includes technical specifications for stability, voltage, and restoration services}
}

@techreport{NESO_Grid_Code_2026,
  author       = {{National Energy System Operator (NESO)}},
  title        = {The Grid Code, Issue 6, Revision 36},
  institution  = {National Energy System Operator},
  address      = {United Kingdom},
  year         = {2026},
  note         = {Full Grid Code}
}

@techreport{NESO_Stability_Y1_Round1_Results_2024,
  author       = {{National Energy System Operator (NESO)}},
  title        = {Stability Mid-Term (Y-1) Year One Results Table},
  institution  = {National Energy System Operator},
  address      = {United Kingdom},
  year         = {2024},
  month        = nov,
  day          = {22},
  url          = {https://www.neso.energy/document/347836/download},
  note         = {Stability Market Tender Results, Round 1}
}

@misc{Netztransparenz2026MarktdesignMomentanreserve,
  author       = {{Netztransparenz} and {Bundesnetzagentur}},
  title        = {Allgemeine Informationen zum Marktdesign -- Information on Market Design},
  year         = {2026},
  month        = jan,
  date         = {2026-01-20},
  howpublished = {Online document},
  url          = {https://www.netztransparenz.de/en/Ancillary-Services/Frequency-stability/Market-based-procurement-of-inertia-of-local-grid-stability},
  language     = {German}
}

@techreport{GermanTSO2025SystemStability,
  title       = {{System Stability Report 2025}},
  author      = {{50Hertz Transmission GmbH} and {Amprion GmbH} and {TenneT TSO GmbH} and {TransnetBW GmbH}},
  year        = {2025},
  month       = {6},
  institution = {50Hertz Transmission GmbH, Amprion GmbH, TenneT TSO GmbH, TransnetBW GmbH},
  url         = {https://www.bundesnetzagentur.de/DE/Fachthemen/ElektrizitaetundGas/NEP/Strom/Systemstabilitaet/start.html},
  note        = {Prepared in accordance with Section 12 i of the Energy Industry Act (EnWG)}
}

@misc{Netztransparenz2026_Verfuegbarkeitsbestimmung,
  author       = {{Netztransparenz.de}},
  title        = {Bestimmung der Verfügbarkeit},
  year         = {2026},
  month        = jan,
  date         = {2026-01-20},
  howpublished = {\url{https://www.netztransparenz.de/en/Ancillary-Services/Frequency-stability/Market-based-procurement-of-inertia-of-local-grid-stability}},
  note         = {Document provided under "Information on availability", Frequency Stability – Market-based procurement of inertia of local grid stability}
}

@misc{Netztransparenz2026_DatenaustauschAbrechnung,
  author       = {{Netztransparenz.de}},
  title        = {Datenaustausch und Abrechnung},
  year         = {2026},
  month        = jan,
  date         = {2026-01-20},
  howpublished = {\url{https://www.netztransparenz.de/en/Ancillary-Services/Frequency-stability/Market-based-procurement-of-inertia-of-local-grid-stability}},
  note         = {Document provided under "Information on data exchange and billing", Frequency Stability – Market-based procurement of inertia of local grid stability}
}

@misc{Netztransparenz2025_AngebotsabgabeAggregation,
  author       = {{Netztransparenz.de}},
  title        = {Rahmenvertragsabschluss und Angebotsabgabe sowie Aggregation von Einheiten},
  year         = {2025},
  month        = oct,
  date         = {2025-10-22},
  howpublished = {\url{https://www.netztransparenz.de/en/Ancillary-Services/Frequency-stability/Market-based-procurement-of-inertia-of-local-grid-stability}},
  note         = {Document provided under "Information on bidding and aggregation", Frequency Stability – Market-based procurement of inertia of local grid stability; slide updates dated 2026-01-20}
}

@techreport{energinet_prequalification_2025,
  title        = {Prequalification of Units and Aggregated Portfolios},
  author       = {{Energinet}},
  institution  = {Energinet},
  year         = {2025},
  month        = {November},
  version      = {2.2.1},
  url          = {https://energinet.dk/el/balancering-og-systemydelser/adgang-til-systemydelsesmarkederne/praekvalifikation-og-test/},
  note         = {Accessed: 2026-03-09}
}

@techreport{ENTSOE2018_LFSM,
  title        = {Limited Frequency Sensitive Mode: ENTSO-E Guidance Document for National Implementation for Network Codes on Grid Connection},
  author       = {{ENTSO-E}},
  institution  = {European Network of Transmission System Operators for Electricity (ENTSO-E)},
  year         = {2018},
  month        = {January},
  address      = {Brussels, Belgium},
  note         = {Published 31 January 2018},
}

@techreport{ENTSOE2018_FSM,
  title        = {Frequency Sensitive Mode: ENTSO-E Guidance Document for National Implementation for Network Codes on Grid Connection},
  author       = {{ENTSO-E}},
  institution  = {European Network of Transmission System Operators for Electricity (ENTSO-E)},
  year         = {2018},
  month        = {January},
  address      = {Brussels, Belgium},
  note         = {Published 31 January 2018},
}

@manual{NESO2026_GridCode,
  title        = {The Grid Code},
  author       = {{National Energy System Operator}},
  organization = {National Energy System Operator (NESO)},
  year         = {2026},
  month        = {April},
  note         = {Issue 6, Revision 37, 13 April 2026},
  address      = {Great Britain},
}

@techreport{Energinet2019_HVDC_Requirements,
  title        = {Requirements for Grid Connection of High-Voltage Direct Current Systems and Direct Current-Connected Power Park Modules (HVDC), Articles 11--54},
  author       = {{Energinet}},
  institution  = {Energinet},
  year         = {2019},
  month        = {October},
  number       = {Doc. 17/15796-41},
  address      = {Fredericia, Denmark},
  note         = {Revision 0, approved by the Danish Utility Regulator; translation version},
}

@techreport{Energinet2025_NC_RfG_V5,
  title        = {NC RfG – National Requirements for Grid Connection of Generation Facilities (Version 5)},
  author       = {{Energinet}},
  institution  = {Energinet},
  year         = {2025},
  month        = {November},
  date         = {2025-11-21},
  number       = {Doc. 25/11731-2},
  address      = {Fredericia, Denmark},
  note         = {Effective from 1 December 2025; English translation for informational purposes (non-legally binding)},
}

@misc{Energinet_survey_nodate,
	title = {Survey on a Potential Split of {FCR}-N},
	url = {https://en.energinet.dk/electricity/balancing-and-ancillary-services/news-about-ancillary-services/2026/03/30/survey-fcr-n/},
    year = {n.d.},
	note = {[accessed: 2026-04-23]},
	langid = {english},
}

@techreport{NESO2026_StaticFFR_Submission,
  author       = {{National Energy System Operator (NESO)}},
  title        = {Static Firm Frequency Response Submission Document: Proposed Changes to the Static Firm Frequency Response Services Terms and Conditions},
  institution  = {National Energy System Operator},
  year         = {2026},
  month        = {March},
  date         = {2026-03-23},
  address      = {Warwick, United Kingdom},
  type         = {Technical Report},
  note         = {Consultation submission document},
}

@techreport{NESO_DynamicResponse_RealTime,
  title        = {Real-time Dynamic Response: Detailed Service Design},
  author       = {{National Energy System Operator (NESO)}},
  institution  = {National Energy System Operator},
  type         = {Technical Report},
  year         = {n.d.},
  note         = {Public document describing real-time Dynamic Response service design}
}

@misc{NESO_ResponseReform_2025_Webinar,
  author       = {{National Energy System Operator (NESO)}},
  title        = {Response Reform Webinar -- January 2025},
  year         = {2025},
  month        = {January},
 url = {https://www.neso.energy/industry-information/balancing-services/frequency-response-services/future-frequency-response},
  note         = {Webinar slides, Public},
}

@misc{EU2017_2195,
  title        = {Commission Regulation (EU) 2017/2195 of 23 November 2017 establishing a guideline on electricity balancing},
  author       = {{European Commission}},
  year         = {2017},
  month        = nov,
  number       = {2017/2195},
  howpublished = {Official Journal of the European Union, L 312},
  pages        = {6--53},
  url          = {https://eur-lex.europa.eu/legal-content/EN/TXT/?uri=CELEX:32017R2195},
  note         = {Accessed: 2026-05-02}
}

@misc{EU2019_943,
  title        = {Regulation (EU) 2019/943 of the European Parliament and of the Council of 5 June 2019 on the internal market for electricity (recast)},
  author       = {{European Parliament and Council of the European Union}},
  year         = {2019},
  month        = jun,
  number       = {2019/943},
  howpublished = {Official Journal of the European Union, L 158},
  pages        = {54--124},
  url          = {https://eur-lex.europa.eu/legal-content/EN/TXT/?uri=CELEX:32019R0943},
  note         = {Accessed: 2026-05-02}
}

@misc{netztransparenz_reactive_power_12h_enwg,
  author       = {{German Transmission System Operators (TSOs)}},
  title        = {Market-based procurement of reactive power in accordance with Section 12h EnWG},
  year         = {n.d.},
  url          = {https://www.netztransparenz.de/en/Ancillary-Services/Voltage-stability/Market-based-procurement-of-reactive-power-in-accordance-with-Section-12h-EnWG},
  note         = {[accessed: 2026-05-05]},
  organization = {Netztransparenz.de}
}

@misc{sse_dogger_bank_reactive_power_2022,
  author       = {{SSE Renewables}},
  title        = {Dogger Bank C in UK offshore wind first to provide reactive power},
  year         = {n.d.},
  month        = feb,
  url          = {https://www.sserenewables.com/news-and-views/2022/02/dogger-bank-c-in-uk-offshore-wind-first-to-provide-reactive-power/},
  note         = {[accessed: 2026-05-05]},
  organization = {SSE Renewables}
}

@misc{ScottishPower2020GlobalFirst,
  author       = {{ScottishPower Renewables}},
  title        = {Global first for ScottishPower as COP countdown starts},
  year         = {2020},
  month        = {November},
  day          = {4},
  url          = {https://www.scottishpowerrenewables.com/w/global-first-for-scottishpower-as-cop-countdown-starts},
  note         = {[accessed: 2026-05-05]}
}

@misc{NextKraftwerke_mFRR,
  author       = {{Next Kraftwerke}},
  title        = {What is mFRR (manual Frequency Restoration Reserve / R3)?},
  year         = {n.d.},
  url          = {https://www.next-kraftwerke.com/knowledge/mfrr},
  note         = {[accessed: 2026-05-06]},
  organization = {Next Kraftwerke}
}

@misc{NextKraftwerke_aFRR,
  author       = {{Next Kraftwerke}},
  title        = {What is aFRR (automatic frequency restoration reserve) and how does it work?},
  year         = {n.d.},
  url          = {https://www.next-kraftwerke.com/knowledge/afrr},
  note         = {[accessed: 2026-05-06]},
  organization = {Next Kraftwerke}
}

@misc{netztransparenz_blackstart_market_procurement,
  title        = {Market-based procurement of black start capability},
  author       = {{German Transmission System Operators}},
  year         = {n.d.},
  url          = {https://www.netztransparenz.de/en/Ancillary-Services/Emergency-restoration/Market-based-procurement-of-black-start-capability},
  urldate      = {2026-05-06},
  organization = {Netztransparenz.de},
  note         = {[accessed: 2026-05-06]}
}

@misc{ACER2024_HVDC_Amendments,
  author       = {{Agency for the Cooperation of Energy Regulators (ACER)}},
  title        = {ACER proposes electricity Grid Connection Network Code amendments to the European Commission},
  year         = {2024},
  month        = dec,
  day          = {20},
  url          = {https://www.acer.europa.eu/news/acer-proposes-electricity-grid-connection-network-code-amendments-european-commission},
  note         = {Accessed: 2026-05-06}
}

@misc{ACER2023_GCNC_Amendments,
  author       = {{Agency for the Cooperation of Energy Regulators (ACER)}},
  title        = {ACER proposes amendments to the electricity grid connection network codes},
  year         = {2023},
  month        = dec,
  day          = {19},
  url          = {https://www.acer.europa.eu/news/acer-proposes-amendments-electricity-grid-connection-network-codes},
  note         = {Accessed: 2026-05-06}
}

@techreport{NGESO_2023_stability_midterm_EOI,
  author       = {{National Grid Electricity System Operator}},
  title        = {Instructions to Tenderers -- EOI and Consultation: Stability Mid-Term Market},
  institution  = {National Grid ESO},
  year         = {2023},
  month        = oct,
  type         = {Tender documentation},
  note         = {Tender Year: 2023--2024; Delivery Year: 2025--2026; Version V1, Initial publication, 3 October 2023},
}

@misc{NESO_future_frequency_response,
  author       = {{National Energy System Operator}},
  title        = {Future of Response Services},
  year         = {2025},
  url          = {https://www.neso.energy/industry-information/balancing-services/frequency-response-services/future-frequency-response},
  note         = {Accessed: 2026-05-06}
}

@misc{neso_transmission_constraint_management,
  author       = {{National Energy System Operator}},
  title        = {Transmission Constraint Management},
  year         = {n.d.},
  url          = {https://www.neso.energy/industry-information/balancing-services/system-security-services/transmission-constraint-management},
  note         = {[accessed: 2026-05-06]},
  organization = {NESO}
}

@misc{netztransparenz_redispatch,
  author       = {{German Transmission System Operators}},
  title        = {Redispatch - Ancillary Services},
  year         = {n.d.},
  url          = {https://www.netztransparenz.de/en/Ancillary-Services/System-operations/Redispatch},
  note         = {[accessed: 2026-05-06]},
  organization = {Netztransparenz.de}
}

@misc{netztransparenz_inertia_local_grid_stability,
  author       = {{German Transmission System Operators}},
  title        = {Market-based Procurement of Inertia of Local Grid Stability},
  year         = {2026},
  url          = {https://www.netztransparenz.de/en/Ancillary-Services/Frequency-stability/Market-based-procurement-of-inertia-of-local-grid-stability},
  note         = {Accessed: 2026-05-06},
  organization = {Netztransparenz.de}
}

@techreport{neso_grid_forming_guidance_2026,
  author       = {{National Energy System Operator}},
  title        = {Grid Forming Guidance Note -- Issue 4},
  institution  = {National Energy System Operator (NESO)},
  year         = {2026},
  month        = May,
  url          = {https://www.neso.energy/document/289921/download},
}

@misc{NESO_DynamicServices_DC_DM_DR,
  author       = {{National Energy System Operator (NESO)}},
  title        = {Dynamic Services (DC/DM/DR)},
  year         = {2026},
  url          = {https://www.neso.energy/industry-information/balancing-services/frequency-response-services/dynamic-services-dcdmdr},
  note         = {Accessed: 2026-05-06},
  organization = {NESO}
}

@misc{ENTSOE_GFM_PhaseII_2025,
  author       = {{ENTSO-E}},
  title        = {ENTSO-E Publishes Phase II Technical Report on Grid Forming Requirements},
  year         = {2025},
  month        = nov,
  day          = {4},
  howpublished = {\url{https://www.entsoe.eu/news/2025/11/04/entso-e-publishes-phase-ii-technical-report-on-grid-forming-requirements/}},
  note         = {Accessed: 2026-05-07}
}

@article{yu_review_2025,
	 author  = {Yu, Yun and Guan, Yajuan and Tarek, Bahaa and Leon, Mitchel Andres Leon and {\"O}zcan, Kemal Onur and Feleke, Solomon and Anteneh, Degarege and Khan, Baseem and Vasquez, Juan C. and Guerrero, Josep M.},
  title   = {A review of international grid codes for wind power integration},
  journal = {Energy Convers Manag X},
  year    = {2025},
  volume  = {28},
  pages   = {101278},
  doi     = {10.1016/j.ecmx.2025.101278},
  url     = {https://www.sciencedirect.com/science/article/pii/S2590174525004106}
}

@ARTICLE{ancillarySurvey_vahid,
  author={Hosseinnezhad, Vahid and Honarmand, Mohammad Esmaeil and Hayes, Barry and Phelan, Michael and Conlon, Paul and Sobral, Paulo Miguel Guilherme Da Costa and Siano, Pierluigi},
  journal={IEEE Access}, 
  title={Ancillary Services in Modern Power Systems: A Practical Survey}, 
  year={2026},
  volume={14},
  number={},
  pages={35608-35632},
  doi={10.1109/ACCESS.2026.3668800}}

@ARTICLE{FreqCtrl_review_IEEEAccess,
  author={Lin, Chung-Han and Wu, Yuan-Kang},
  journal={IEEE Access}, 
  title={Overview of Frequency-Control Technologies for a VSC-HVDC-Integrated Wind Farm}, 
  year={2021},
  volume={9},
  number={},
  pages={112893-112921},
  doi={10.1109/ACCESS.2021.3102829}}

@article{Ullah2024WindStorageFR,
  title = {A comprehensive review of wind power integration and energy storage technologies for modern grid frequency regulation},
  author = {Ullah, Farhan and Zhang, Xuexia and Khan, Mansoor and Mastoi, Muhammad Shahid and Munir, Hafiz Mudassir and Flah, Aymen and Said, Yahia},
  journal = {Heliyon},
  volume = {10},
  number = {},
  pages = {e30466},
  year = {2024},
  publisher = {Elsevier},
  issn = {2405-8440},
  doi = {10.1016/j.heliyon.2024.e30466},
  url = {https://doi.org/10.1016/j.heliyon.2024.e30466}
}

@article{Boyle2024FrequencyControlWindIreland,
  title = {A review of frequency-control techniques for wind power stations to enable higher penetration of renewables onto the Irish power system},
  author = {Boyle, James and Littler, Timothy},
  journal = {Energy Reports},
  volume = {12},
  pages = {5567--5581},
  year = {2024},
  publisher = {Elsevier},
  issn = {2352-4847},
  doi = {10.1016/j.egyr.2024.11.031},
  url = {https://doi.org/10.1016/j.egyr.2024.11.031}
}

@INPROCEEDINGS{FreqCtrlReview_Conf,
  author={Wang, Zhuzhu and Wu, Lei},
  booktitle={2024 56th North American Power Symposium (NAPS)}, 
  title={A Review on Control Schemes of HVDC-Integrated Offshore Wind Farms for Fast Frequency Support}, 
  year={2024},
  volume={},
  number={},
  pages={1-6},
  doi={10.1109/NAPS61145.2024.10741763}}

@misc{netztransparenz_inertia_2026,
  author       = {{German Transmission System Operators}},
  title        = {Market-based Procurement of Inertia of Local Grid Stability},
  year         = {2026},
  url          = {https://www.netztransparenz.de/en/Ancillary-Services/Frequency-stability/Market-based-procurement-of-inertia-of-local-grid-stability},
  organization = {Netztransparenz.de},
  urldate      = {2026-05-11}
}

@patent{bjorn2022power,
  title        = {Power Ramp Rate Control},
  author       = {Bjorn, Hans Kristian and Knudsen, Jan Vestergaard and Saibabu, Sanka},
  number       = {US 11525433 B2},
  type         = {Patent},
  nationality  = {US},
  assignee     = {Vestas Wind Systems A/S},
  year         = {2022},
  month        = dec,
  day          = {13},
  filingdate   = {2018-11-20},
  prioritydate = {2017-12-21},
  note         = {U.S. Patent No. 11,525,433 B2}
}

@patent{heidenreich2015wind,
  title        = {Wind Turbine Torque Limiting Clutch System},
  author       = {Heidenreich, David C. and Cole, Richard E. Jr.},
  number       = {US 9097239 B2},
  type         = {Patent},
  nationality  = {US},
  assignee     = {EBO Group, Inc.},
  year         = {2015},
  month        = aug,
  day          = {4},
  filingdate   = {2012-04-24},
  note         = {U.S. Patent No. 9,097,239 B2}
}

@techreport{Energinet2026EvidenceStabilityMarket,
  author      = {{Energinet System Stability}},
  title       = {{From Evidence to Stability Market: The Role of Grid Forming Capability and Performance in the Danish Power System}},
  institution = {{Energinet}},
  type        = {Report},
  number      = {Doc. 20/00794-25},
  address     = {Fredericia, Denmark},
  year        = {2026},
  month       = may,
  note        = {Offentlig/Public}
}

@inproceedings{Heid2022Asymmetric,
  author    = {Heid, Johannes and Schittek, Walter and Hachmann, Christian and Braun, Martin},
  title     = {Asymmetric Contributions to Instantaneous Reserve by Generation, Loads, and Storage},
  booktitle = {Tagung Zuk{\"u}nftige Stromnetze 2022},
  year      = {2022},
  month     = jan,
  doi       = {10.17170/kobra-202202015687},
  note      = {Postprint}
}

@ARTICLE{10981620,
  author={Salem, Qusay and Fawaz, Bayan Bany and Aljarrah, Rafat and Karimi, Mazaher},
  journal={IEEE Open Journal of the Industrial Electronics Society}, 
  title={Grid Forming Converters for Low Inertia Systems-Capabilities and Limitations: A Critical Review}, 
  year={2025},
  volume={6},
  number={},
  pages={775-801},
  doi={10.1109/OJIES.2025.3566213}}

@ARTICLE{11540169,
  author={Schittek, Walter and Hachmann, Christian and Wiese, Nils and Prieto-Araujo, Eduardo and Brammer, Sofie and Ferry, Jurian and Gomis-Bellmunt, Oriol and Braun, Martin},
  journal={IEEE Open Access Journal of Power and Energy}, 
  title={Complementary Grid Forming: Inertia Gathered from Constrained Resources}, 
  year={2026},
  volume={},
  number={},
  pages={1-1},
  doi={10.1109/OAJPE.2026.3698146}}

@article{Rehman2026Asymmetric,
  author  = {Rehman, Syed Muhammad Sami ur and Lens, Hendrik},
  title   = {Asymmetric Inertia Provision to Power Systems by Grid-Forming Inverter-Based Resources: Modeling, Control, and Stability Implications},
  journal = {Preprint submitted to Elsevier},
  year    = {2026},
  note    = {Preprint}
}

@misc{EnerginetPricesProcurementProjections,
  author       = {{Energinet}},
  title        = {{Prices, Procurement and Projections}},
  year = {n.d.},
  howpublished = {\url{https://en.energinet.dk/electricity/balancing-and-ancillary-services/prices-procurement-and-projections/}},
  note         = {accessed: 2026-07-05}
}

@article{saborio2020communication,
  title={Communication-less frequency support from offshore wind farms connected to HVDC via diode rectifiers},
  author={Sabor{\'\i}o-Romano, Oscar and Bidadfar, Ali and Sakamuri, Jayachandra N and Zeni, Lorenzo and G{\"o}ksu, {\"O}mer and Cutululis, Nicolaos A},
  journal={IEEE Transactions on Sustainable Energy},
  volume={12},
  number={1},
  pages={441--450},
  year={2020},
  publisher={IEEE}
}

@article{jiang2023novel,
  title={A novel coordinated control strategy for frequency regulation of MMC-HVDC connecting offshore wind farms},
  author={Jiang, Shouqi and Wang, Hanbing and Li, Guoqing and Xin, Yechun and Wang, Lixin and Xu, Yanan},
  journal={IEEE Transactions on Sustainable Energy},
  volume={15},
  number={2},
  pages={1028--1038},
  year={2023},
  publisher={IEEE}
}

@article{Xu2026HolisticGFM,
  author        = {Xu, Zhenghua and Gro{\ss}, Dominic and Raducu, George Alexandru and Nouri, Banafsheh and Sabor{\'i}o-Romano, Oscar and Cutululis, Nicolaos Antonio},
  title         = {Holistic Grid-Forming Control to Enhance the Frequency Support from {HVDC}-Connected Offshore Wind Power Plants},
  journal       = {arXiv preprint arXiv:2605.23041},
  year          = {2026},
  month         = may,
  eprint        = {2605.23041},
  archivePrefix = {arXiv},
  primaryClass  = {eess.SY}
}

@techreport{Kwon2026DeploymentGFM,
  author      = {Kwon, Jun Bum and Liao, Yicheng and Lu, Liang and Gong, Hong and Bose, Anurag},
  title       = {Deployment of Grid Forming Technology in the Danish Power System: Technical Report Version 2},
  institution = {Energinet},
  type        = {Technical Report},
  number      = {Doc. 20/00794-25},
  year        = {2026},
  month       = may,
  note        = {May 15, 2026},
  pages       = {110}
}

@ARTICLE{10158921,
  author={Yan, Kefei and Li, Guoqing and Zhang, Rufeng and Xu, Yan and Jiang, Tao and Li, Xue},
  journal={IEEE Transactions on Power Systems}, 
  title={Frequency Control and Optimal Operation of Low-Inertia Power Systems With HVDC and Renewable Energy: A Review}, 
  year={2024},
  volume={39},
  number={2},
  pages={4279-4295},
  doi={10.1109/TPWRS.2023.3288086}}

@ARTICLE{10130020,
  author={Liu, Haoyu and Liu, Chongru},
  journal={Journal of Modern Power Systems and Clean Energy}, 
  title={Frequency Regulation of VSC-MTDC System with Offshore Wind Farms}, 
  year={2024},
  volume={12},
  number={1},
  pages={275-286},
  doi={10.35833/MPCE.2023.000001}}





\end{document}